\documentclass[aps,prb,twocolumn]{revtex4}
\usepackage{graphics,epsfig,amsfonts,amssymb,amsmath}
\begin{document}
\title{Optical response for the d-density wave model}

\author{B. Valenzuela$^1$, E.J. Nicol$^1$, J.P. Carbotte$^2$}
\affiliation{$^1$Department of Physics, University of Guelph, Ontario, 
N1G 2W1, Canada}
\affiliation{$^2$Department of Physics and Astronomy, McMaster University, 
Hamilton, Ontario, L8S 4M1, Canada}

\begin{abstract}
We have calculated the optical conductivity and the Raman response for 
the d-density wave model, proposed as a possible explanation for the 
pseudogap seen in high $T_c$
cuprates. The total optical spectral weight remains 
approximately constant on opening of the pseudogap at fixed temperature. 
This occurs 
because there is a transfer of weight from the Drude peak to interband 
transitions across the pseudogap. 
The interband peak in the optical conductivity is 
prominent but becomes progressively reduced 
with increasing temperature, 
with impurity scattering,
which distributes it over a larger energy range,
and with inelastic scattering which can also shift its
position, making it difficult to have a direct determination of the value of the
pseudogap. Corresponding structure 
is seen in the optical scattering rate, but not necessarily at the
same energies as in the conductivity.
\end{abstract}

\date{\today}
\maketitle
\section{Introduction}

The pseudogap in the cuprates is widely believed to hold
an important clue to the fundamental nature of these 
systems.\cite{timuskpseudo}
Through angular
resolved photoemission studies\cite{ding,shen} of the variation around
the Fermi surface of the leading edge of the electron spectral
density, it has been established that the pseudogap has d-wave symmetry
as does the superconducting gap. A smooth evolution
of the superconducting gap into the pseudogap is also
observed in tunneling\cite{renner} which shows that, in the
underdoped regime, the gap fills in with increasing temperature
but does not close as one goes through $T_c$.
These ideas have led to a
school of thought that believes that the pseudogap is a
precursor to superconductivity.
One model is the 
preformed pair\cite{kivelson,emery} model with phase coherence lost at the 
superconducting $T_c$, but the pairs remain till the higher 
pseudogap temperature $T^*$. Another model includes finite 
momentum pairs\cite{chen}, which exist at any finite temperature below 
$T^*$ but do not contribute to the superfluid stiffness. No zero 
momentum Cooper pairs remain above $T_c$ where superconductivity is lost.

A second school of thought envisages that the pseudogap
has its origin in a competing order parameter.\cite{chakra} The DDW theory 
is found among this group. It establishes a charge density wave order 
which has d-wave
symmetry and which opens at the antiferromagnetic Brillouin zone as 
the competing order parameter.
The DDW model\cite{chakra,chakra2,wang,carbotte,nayak} breaks time reversal
symmetry because it introduces bond currents which also have
small orbital magnetic moments. The two dimensional $CuO_2$ Brillouin 
zone is halved into a lower and an upper antiferromagnetic Brillouin zone.
The DDW model can be included in the group of models associated with  
unconventional density waves (UDW), which have been 
proposed to explain some phases of correlated electron systems.
For instance, UDWs have been 
used to understand 
quasi-two dimensional organic conductors, such as
$\alpha-(BEDT-TTF)_2 MHg(SCN)_4$, with $M=K, Rb$ and $Tl$.\cite{maki,dora}
Unconventional density waves have a gap which averages to zero 
on the 
Fermi surface (although in contrast, the DDW gap,
not opening at the Fermi surface, can only average to
zero over the Brillouin zone). 
Likewise, there is no periodic modulation 
in the charge or spin density, which makes them difficult to detect 
experimentally. For this reason it is very important to have
predictions for transport properties in order to look for 
features that may characterize them.

In this paper, we calculate the
optical conductivity and the electronic Raman response for the DDW model. 
For tight-binding bands with first-nearest-neighbours only, the chemical potential 
acts as the lower cutoff on the 
interband signal in both optical responses as first discussed by 
Yang and Nayak\cite{nayak} in the case of optical conductivity. As 
the second-nearest-neighbour hopping is increased from zero, 
the cutoff remains, but shifts and is no longer at 
twice the chemical potential. The 
role of the chemical 
potential in the optical response is intimately related to 
the fact that the DDW order parameter opens at 
the antiferromagnetic Brillouin zone.
As a consequence optical spectral weight is transferred from 
intraband to interband transitions in the energy range of the gap with 
little change in total spectral weight. 
We obtain a 
peak in the real part of the conductivity coming from 
the interband contribution. This feature is 
robust with respect to different dispersion relations with 
second-nearest-neighbours. It is modified by
impurities and by inelastic scattering, which shift
the energy of the structure and obscure
the cutoff, making an unambiguous determination of the pseudogap
difficult. As
the temperature is increased, the interband spectral weight 
loses intensity. It is fully depleted at the 
pseudogap temperature $T^*$ at which point it has all been 
transferred to the intraband.
A similar peak is found in the optical scattering rate, but it 
can be shifted in energy. 
This feature can be used to test for 
the DDW state, but so far, it has not been 
seen in the cuprates.

The paper is organized as follows. In the next section, 
we present the DDW model, then in section \ref{sec:one-part}, 
we discuss the one-particle 
properties relevant for our problem. 
Section \ref{sec:conduc} treats the problem of the 
optical conductivity and of the optical spectral weight. 
Section \ref{sec:raman} 
deals with the electronic 
Raman response. In section \ref{sec:scatt}, 
we consider the effects of elastic and inelastic 
scattering, and in the final section our 
conclusions are presented.

\section{Hamiltonian}
We consider a two dimensional system of electrons with the Hamiltonian given by:
\begin{eqnarray}
\hat{H}&=&
\sum_{{\bf k}\sigma} (\epsilon_{\bf k}-\mu) c^\dag_{{\bf k}\sigma} c_{{\bf k}\sigma} 
\nonumber\\{}&&+\frac{1}{2}\sum_{\substack{{\bf k}{\bf k'}\sigma\sigma' \\ 
{\bf q}}}
V({\bf k},{\bf k'},{\bf q}) 
c_{{\bf {k+q}}\sigma}^\dag c_{{\bf {k'-q}}\sigma'}^\dag 
c_{{\bf k'}\sigma'} c_{{\bf k}\sigma},
\label{eq:Ham}
\end{eqnarray}
where $c^\dag_{{\bf k},\sigma}$ and $c_{{\bf k},\sigma}$ are the creation and 
annihilation operators of electrons with momentum ${\bf k}$ 
over the entire Brillouin zone of the $Cu O_2$ plane (2D) and 
spin $\sigma$, $\mu$ is the chemical potential, $\epsilon_{\bf k}$ is the 
dispersion relation and $V({\bf k,k',q})$ is the 
interaction between the 
electrons. Following D\'ora {\it et al.}\cite{dora}, the interaction is taken
to be of the form 
$V({\bf k,k',q})=2Vf({\bf k})f({\bf k'})\delta({\bf q-Q})$, where 
${\bf Q}=(\pi,\pi)$ is the wavevector where the density wave order 
takes place. 

Using the definition of the density wave order 
parameter,
\begin{equation}
i\Delta_{\bf k}=-Vf({\bf k})\sum_{{\bf k'}\sigma}f({\bf k'})
\langle c_{{\bf {k'+Q}}\sigma}^\dag c_{{\bf k'}\sigma}\rangle,
\end{equation}
then, to within an additive constant, one arrives at the 
following effective Hamiltonian:  
\begin{eqnarray}
\hat{H}_{DDW}=
\sum_{{\bf k}\sigma} \Big[(\epsilon_{\bf k}-\mu) c^\dag_{{\bf k}\sigma} c_{{\bf k}\sigma} 
+i\Delta_{\bf k} c^\dag_{{\bf k}\sigma} c_{{\bf k+Q}\sigma}\Big].
\label{eq:dDW-H}
\end{eqnarray}
To diagonalize the Hamiltonian in
Eq.~(\ref{eq:dDW-H}), we write it with momentum restricted to 
the first antiferromagnetic Brillouin zone using 
Nambu notation:
\begin{equation}
\label{eq:nambu-H}
\hat{H}_{DDW}=\sum_{{\bf k},\sigma} \Psi_\sigma^\dag({\bf k})
\Big(\epsilon_{\bf k}^-\hat{\tau}_3+
(\epsilon_{\bf k}^+-\mu)\hat{\tau}_0-\Delta_{\bf k}\hat{\tau}_2\Big)
\Psi_\sigma({\bf k}),
\end{equation}
where $\Psi_\sigma^\dag({\bf k})=(c^\dag_{{\bf k},\sigma},
c^\dag_{{\bf k+Q},\sigma})$, $\hat{\tau}_i$ stands for 
the Pauli matrices, 
$\epsilon_{\bf k}^-=(\epsilon_{\bf k}-\epsilon_{\bf k-Q})/2$ is 
the nesting dispersion relation and 
$\epsilon_{\bf k}^+=(\epsilon_{\bf k}+\epsilon_{\bf k-Q})/2$ is the 
imperfect nesting one.
 
Using a standard Bogoliubov transformation, the coherence 
factors are found to be:
\begin{eqnarray}
\label{eq:coherent}
u_{\bf k}=\sqrt{\frac{1}{2}\Big(1+ \frac{\epsilon_{\bf k}^-}
{E_{\bf k}}\Big)},\nonumber\\
v_{\bf k}=\sqrt{\frac{1}{2}\Big(1- \frac{\epsilon_{\bf k}^-}{E_{\bf k}}
\Big)},
\end{eqnarray}
where $E_{\bf k}=\sqrt{(\epsilon_{\bf k}^-)^2 + \Delta_{\bf k}^2}$. 
The energy eigenvalues are:
\begin{equation}
\label{eq:bands}
E_{\bf k}^\pm=\epsilon_{\bf k}^+ \pm E_{\bf k},
\end{equation}
where $\pm$ refers to the upper and the lower antiferromagnetic Brillouin zone, 
respectively.
The imperfect nesting dispersion
$\epsilon_{\bf k}^+$ does not enter into the coherence factors
in Eq.~(\ref{eq:coherent}), as has been previously discussed\cite{chakraphoto},
however, the effects induced by the imperfect nesting term enter
  in Eq.~(\ref{eq:bands}), where the it is outside the square-root. 
As a consequence, in all physical quantities, this term enters 
linearly along with the chemical potential. 
This fact will be of importance for the 
optical response. 

From the Hamiltonian of Eq.~(\ref{eq:nambu-H}), we can write straightforwardly
the Green's function
in Nambu notation,
$\hat{G}({\bf k},i\omega_n)^{-1}=i\omega_n-\hat{H}_{DDW}$, as follows:
\begin{equation}
\label{eq:greenfun}
\hat{G}({\bf k},i\omega_n)^{-1}=(i\omega_n + \mu-\epsilon_{\bf k}^+)
\hat{\tau}_0-\epsilon_{\bf k}^-\hat{\tau}_3+\Delta_{\bf k}
\hat{\tau}_2,
\end{equation}
where the $\omega_n$ are the fermion Matsubara frequencies.
The spectral functions
$A_{ij}({\bf k},\omega)=-2 \rm Im G_{ij}({\bf k},\omega+i\delta)$ are the following:
\begin{eqnarray}
\label{eq:A}
A_{11}({\bf k},\omega)&=&2\pi\big\{u_{\bf k}^2\delta(\omega+\mu-E_{\bf k}^+)
+v_{\bf k}^2\delta(\omega+\mu-E_{\bf k}^-)\big\}, \nonumber\\
A_{22}({\bf k},\omega)&=&2\pi\big\{v_{\bf k}^2\delta(\omega+\mu-E_{\bf k}^+)
+u_{\bf k}^2\delta(\omega+\mu-E_{\bf k}^-)\big\},\nonumber\\
A_{12}({\bf k},\omega)&=&A_{21}({\bf k},\omega)\nonumber\\
&=&2\pi v_{\bf k}u_{\bf k}
\big\{\delta(\omega+\mu-E_{\bf k}^+)-\delta(\omega+\mu-E_{\bf k}^-)
\big\},\nonumber\\
\end{eqnarray}
We observe that each $A_{ij}$ depends on the upper and lower bands,
which reflects the physics of DDW theory
where the quasiparticles spend time in both bands ({\it i.e.}, the
 upper as well 
as the lower antiferromagnetic Brillouin zone).

We study the case  with 
 $\Delta_{\bf k}=\frac{\Delta}{2}(\cos(k_x)-\cos(k_y))$ 
and the dispersion relation is of the form $\epsilon_{\bf k}=\epsilon_{\bf k}^-
+\epsilon_{\bf k}^+$, where 
$\epsilon_{\bf k}^-=-2t(\cos k_x+\cos k_y)$ and
$\epsilon_{\bf k}^+=-4t'\cos k_x\cos k_y$. We choose $t'<0$, which is
the relevant case for the 
cuprates. All the quantities are in units of 
the nearest-neighbour hopping parameter $t$, 
 typically chosen to be between $0.1-0.25$eV. 
Here, $t'$ is the next-nearest-neighbour 
hopping parameter.

\section{Bands and density of states}
\label{sec:one-part}
\begin{figure}
\includegraphics[clip,width=0.45\textwidth]{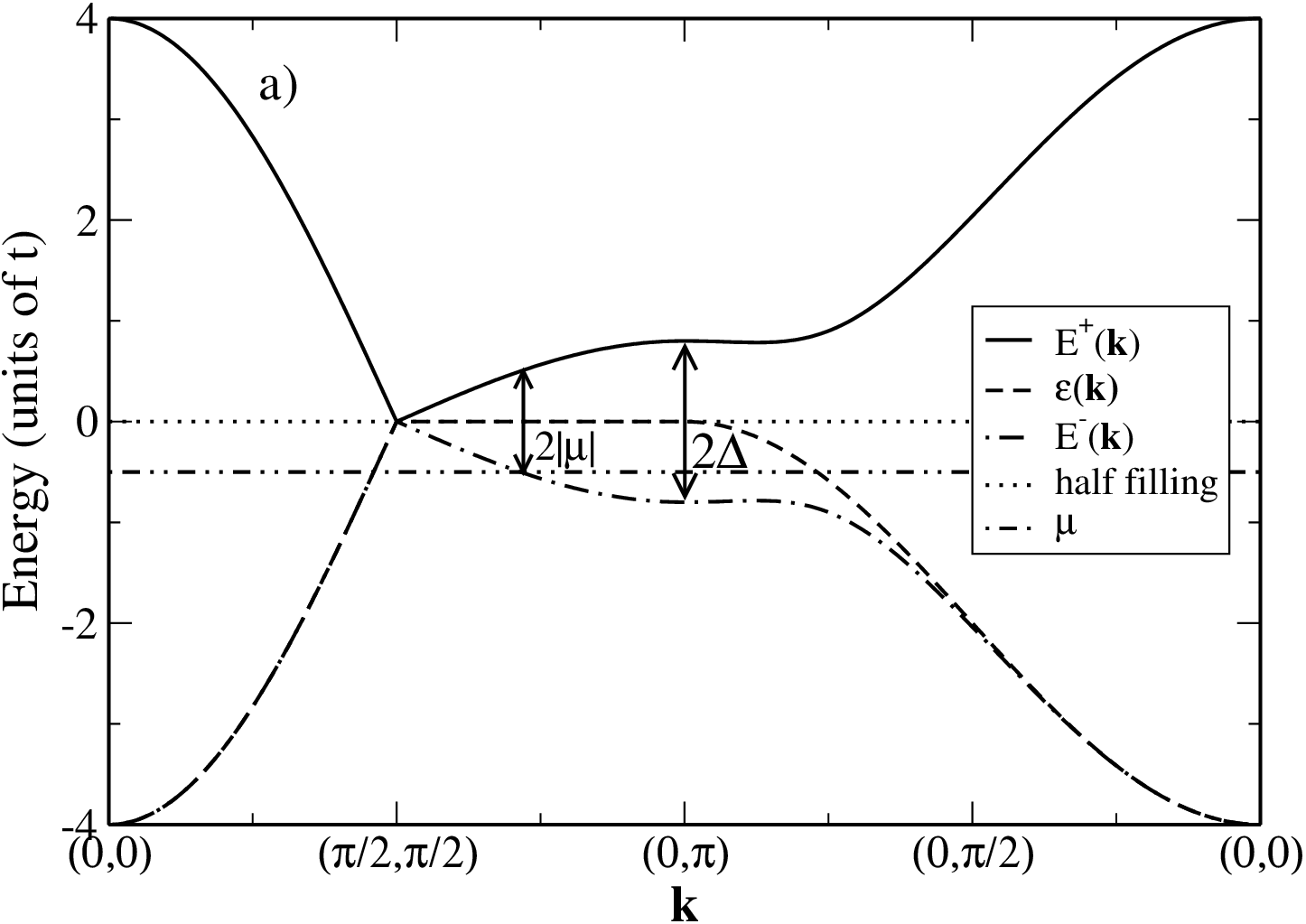}
\includegraphics[clip,width=0.45\textwidth]{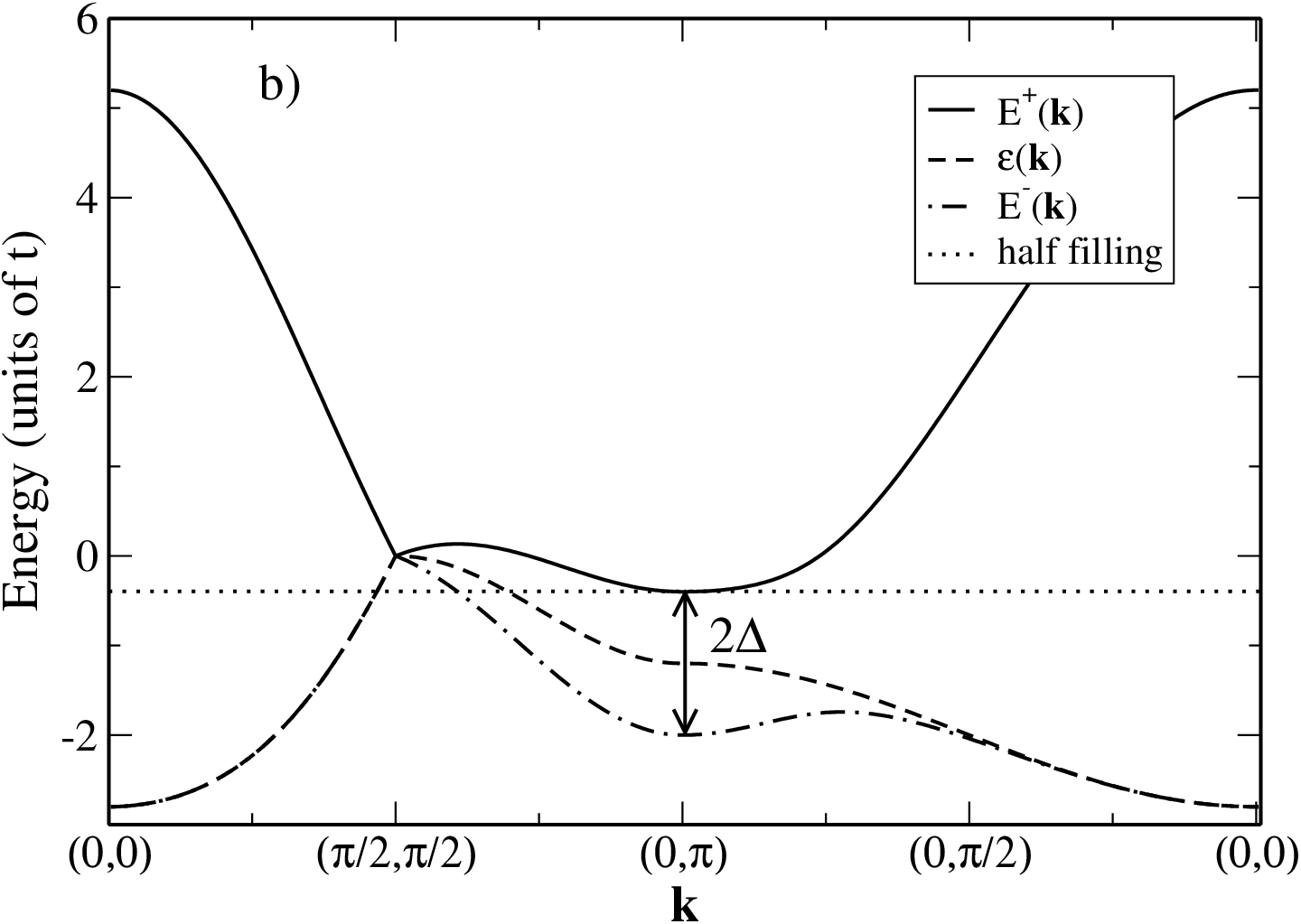}
\caption{Energy dispersions $E_{\bf}^+$ and $E_{\bf}^-$ of the DDW model 
for (a) $t'=0$  and
(b) $t'=-0.3t$, in the case of 
$\Delta=0.8t$. The horizontal lines indicate half filling (dotted)
and a chemical potential $2|\mu|=1.0t$ (dash-double-dotted, shown only
in (a)). 
The tight-binding energy dispersion 
(dashed line) is presented for comparison.} 
\label{fig:bands}
\end{figure}

\begin{figure}
\includegraphics[clip,width=0.45\textwidth]{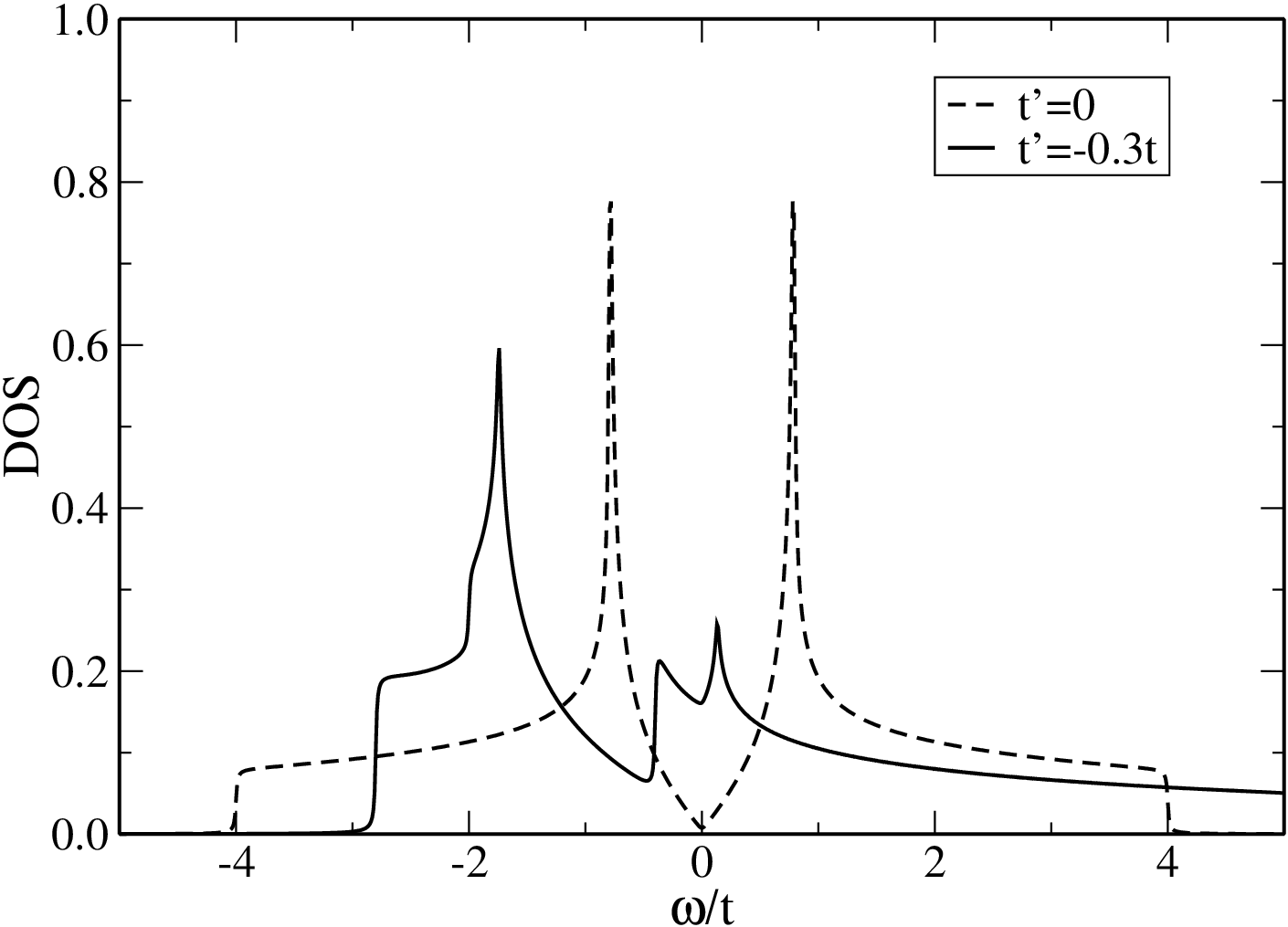}
\caption{Density of states for a DDW system with $\Delta=0.8t$,
for $t'=0$ and $t'=-0.3t$. 
The lower band narrows with increasing $t'$ and the density of states
acquires significant finite weight within the extended band. 
}
\label{fig:dos}
\end{figure}

In the following we will comment on the band structure given by 
Eq.~(\ref{eq:bands}) together with the density of states. These 
one-particle properties can facilitate our understanding of the optical 
response. 
In Fig.~\ref{fig:bands}, we show the bands given by Eq.~(\ref{eq:bands})
and in Fig.~\ref{fig:dos} we plot the corresponding density of states given 
by:
\begin{equation}
\label{eq:dos}
\rho(\omega)=\sum_{\bf k} \Big[u_{\bf k}^2 \delta(\omega-E_{\bf k}^+)
+v_{\bf k}^2 \delta(\omega-E_{\bf k}^-)\Big],
\end{equation}
where the momentum sum is over the entire Brillouin zone, and this
quantity does not depend on the chemical potential as it can be absorbed
into the frequency.
First, consider the case where 
$t'=0$.
The upper, $E_{\bf k}^+$, and lower, $E_{\bf k}^-$,
bands are shown in Fig.~\ref{fig:bands}a
for $\Delta=0.8t$, 
illustrating that the gap is maximum at $(\pi,0)$ and 
closes at $(\pi/2,\pi/2)$. For this band structure, at
each value of momentum there are two branches, one for 
the upper and one for the lower antiferromagnetic Brillouin zone.
This allows for vertical transitions with zero momentum transfer 
to be induced by the photon field from the lower to the upper band
(interband transitions). 
Indicating schematically a Fermi level $\mu$ on the
band structure diagram in
Fig.~\ref{fig:bands}a, such a transition from the Fermi level
at $\mu$ in the lower band (hole doping) to 
the upper band, by an energy equal to $2|\mu|$, is shown. 
The significance of this result is that, in the absence of other mechanisms for
absorption, such as impurity- or boson-assisted scattering, the minimum energy
for an optical transition at finite frequency is $2|\mu|$.
This $2|\mu|$ becomes the cutoff of the interband optical 
signal at low frequencies 
as will be seen from our full calculations.
The corresponding density of states for this bandstructure
is shown in Fig.~\ref{fig:dos} as 
the dashed line, which
is the conventional one for a d-wave order parameter. 
It has zero states at $\omega=0$. 
For this case of $t'=0$, an analytical expression is available 
for the density of states in the
continuum limit: 
$\frac{\rho(\omega)}{N(0)}=\frac{4}{\pi}K(\frac{\Delta^2}{\omega^2})$,
where $K(x^2)$ is the complete Elliptic integral of
the first kind and $N(0)$ is the density of states at the Fermi level
in the state without the pseudogap.
With $t'$ nonzero, the bands
and the
density of states are modified. For small $t'$ ($|t'| < \Delta/4$), the band 
structure is similar to the one at $t'=0$ 
with the lower band narrower and 
the upper band wider. 
If, on the other hand, $t'$ is big ($|t'| > \Delta/4$), 
the bands overlap, as
 shown in Fig.~\ref{fig:bands}b for $t'=-0.3t$
and the density of states is finite everywhere within the extended band.

\section{Optical conductivity}
\label{sec:conduc}
We consider now the real part of the optical conductivity 
$\sigma(\Omega)$. In the long wavelength limit (${\bf q}\to 0$) the current 
operator in momentum space is:
\begin{eqnarray}
{\bf j}({\bf q}\to 0,i\Omega_n)&=&-eT \sum_{\bf k}\sum_{\omega_m} \Psi_\sigma^\dag({\bf k},
i\omega_m+i\Omega_n) 
(\nabla\epsilon_{\bf k}^+\hat{\tau}_0\nonumber\\
&+&\nabla\epsilon_{\bf k}^-\hat{\tau}_3)
\Psi_\sigma({\bf k},i\omega_m),
\end{eqnarray}
where the $\Omega_n$ are the boson Matsubara frequencies. 
Within linear response theory, the optical conductivity 
is obtained from the current-current 
correlation function $\Pi({\bf q}\to 0,i\Omega_n)$ as 
${\rm Re}\{\sigma_{\alpha\alpha}(\Omega)\}=-(1/\Omega)
\rm Im\{\Pi_{\alpha\alpha}(i\Omega_n\to\Omega+i\delta)\}$, where
$\alpha$ is a Cartesian coordinate, $x,y,z$.
 On the imaginary axis, $\Pi_{\alpha\alpha}(i\Omega_n)$ is given 
in the bubble approximation in terms of the Matsubara Green's functions 
$\hat{G}({\bf k},i\omega_m)$ by the following equation:
\begin{eqnarray}
\Pi(i\Omega_n)&=&e^2T\sum_{{\bf k},\omega_m}{\rm Tr}
\Big[\hat{G}({\bf k},i\omega_m+i\Omega_n)
(\nabla\epsilon_{\bf k}^+\hat{\tau}_0+\nabla\epsilon_{\bf k}^-\hat{\tau}_3)
\nonumber\\ {}&&
\times\hat{G}({\bf k},i\omega_m)
(\nabla\epsilon_{\bf k}^+\hat{\tau}_0+\nabla\epsilon_{\bf k}^-\hat{\tau}_3)
\Big].
\end{eqnarray}
When the Green's function, given by 
Eq.~(\ref{eq:greenfun}), is expressed in terms of  
the spectral functions $A_{ij}({\bf k},\omega)$, the real part of 
the optical conductivity takes the form:
\begin{eqnarray}
\label{eq:resigma}
\lefteqn{\rm Re\{\sigma_{xx}(\Omega)\}=e^2\int^{\infty}_{-\infty} \frac{d\omega}{2\pi}
\frac{f(\omega)-f(\omega+\Omega)}{\Omega}
\times{}} \nonumber\\ &&{}
\sum_{\bf k}\Big\{\Big[\Big(\frac{\partial\epsilon_{\bf k}^+}{\partial k_x}\Big)^2+
\Big(\frac{\partial\epsilon_{\bf k}^-}{\partial k_x}\Big)^2\Big]A_{11}({\bf k},\omega+\Omega)
A_{11}({\bf k},\omega)+\nonumber\\ &&{}
\Big[\Big(\frac{\partial\epsilon_{\bf k}^+}{\partial k_x}\Big)^2-
\Big(\frac{\partial\epsilon_{\bf k}^-}{\partial k_x}\Big)^2\Big]
A_{12}({\bf k},\omega+\Omega)A_{21}({\bf k},\omega)\Big\},
\end{eqnarray}
where the Fermi function is given as $f(\omega)=1/({\rm exp}(\beta\omega)+1)$
and $\beta=1/k_BT$.
If we do not include disorder in the system, we 
can express the conductivity as 
${\rm Re} \{\sigma(\Omega)\}=\sigma_0\delta(\Omega)+\sigma_{inter}(\Omega)$,
where $\sigma_0$ indicates the intraband conductivity 
or Drude weight and 
$\sigma_{inter}(\Omega)$, the interband conductivity.
Replacing the $A_{ij}$ by their expressions in (\ref{eq:A}), we obtain for 
the weight $\sigma_0$ of the intraband conductivity $\sigma_{intra}=\sigma_0\delta(\Omega)$:
\begin{eqnarray}
\label{eq:intra}
\sigma_0&=&2\pi e^2\int_{-\infty}^{\infty} d\omega 
\biggl(-\frac{\partial f(\omega)}{\partial \omega}\biggr)\times
\nonumber\\&&
\sum_{\bf k} 
\Big\{\Big[
\Big(\frac{\partial\epsilon^+_{\bf k}}{\partial k_x}\Big)^2
+\frac{\epsilon^-_{\bf k}}{E_{\bf k}}\Big(\frac{\partial\epsilon^-_{\bf k}}{\partial k_x}\Big)^2\Big]
u_{\bf k}^2\delta(\omega+\mu-E_{\bf k}^+)\nonumber\\
&+&\Big[\Big(\frac{\partial\epsilon^+_{\bf k}}{\partial k_x}\Big)^2-\frac{\epsilon^-_{\bf k}}{E_{\bf k}}\Big(\frac{\partial\epsilon^-_{\bf k}}{\partial k_x}\Big)^2\Big]v_{\bf k}^2\delta(\omega+\mu-E_{\bf k}^-)\Big\},
\end{eqnarray}
and for the interband conductivity for $\Omega>0$:
\begin{eqnarray}
\label{eq:inter}
\sigma_{inter}(\Omega)&=&\frac{\pi e^2}{\Omega}\sum_{\bf k}
\frac{\sinh(\beta\Omega/2)}{\cosh\beta(\epsilon_{\bf k}^+-\mu)+
\cosh(\beta\Omega/2)}\times \nonumber\\{}&&
\Bigl(\frac{\partial\epsilon_{\bf k}^-}{\partial k_x}\Bigr)^2\frac{\Delta_{\bf k}^2}{E_{\bf k}^2}
\delta(\Omega-2E_{\bf k}).
\end{eqnarray}
Concentrating first on the intraband conductivity given by
Eq.~(\ref{eq:intra}), 
if we take both $t'=0$ and the continuum limit, we can extract
analytical information. For $t'=0$, Eq.~(\ref{eq:intra}) reduces to: 
\begin{equation}
\label{eq:sigma0}
\frac{\sigma_0}{\pi e^2}=\int_{-\infty}^{\infty} d\omega 
\biggl(-\frac{\partial f(\omega)}{\partial \omega}\biggr) g_1(\omega),
\end{equation}
where
\begin{equation}
g_1(\omega)=\sum_{\bf k} 
\Big(\frac{\partial \epsilon_{\bf k}}{\partial k_x}\Big)^2
\frac{ \epsilon_{\bf k}^2}{E_{\bf k}^2}\Big(\delta(\omega+\mu-E_{\bf k})
+\delta(\omega+\mu+E_{\bf k})\Big),
\label{eq:g1}
\end{equation}
which can be further reduced in the continuum limit, after integration 
over energy, to:
\begin{equation}
g_1(\omega)=(\hbar v_{Fx})^2 N(0) \frac{4}{\pi}\int_0^{\pi/2}{d\theta}\,
{\rm Re}\left\{\sqrt{1-\frac{\Delta^2}{(\omega+\mu)^2}
\cos^2\theta}
\right\}
\label{eq:g1con}
\end{equation}
where we have made the angular dependence of $\Delta_\theta$ explicit in
Eq.~(\ref{eq:g1con}). Using the result that in
 two dimensions $v_{Fx}^2=v_F^2/2$, where 
$v_F$ is the Fermi velocity, and setting $\hbar$ to 1, 
Eq.~(\ref{eq:g1con}) can 
be finally stated in terms of Elliptic functions as follows:
\begin{eqnarray}
\frac{g_1(\omega)}{2N(0)v_F^2/\pi}
&=&E\Big(\frac{\Delta^2}
{(\omega+\mu)^2}\Big),
 {\rm for} \frac{\Delta}{|\omega+\mu|}<1 \nonumber\\
&=&\Big\{\Big(\frac{|\omega+\mu|}{\Delta}-
\frac{\Delta}{|\omega+\mu|}\Big)K\Big(\frac{(\omega+\mu)^2}
{\Delta^2}\Big)
\nonumber\\{}&&
+\frac{\Delta}{|\omega+\mu|}E\Big(\frac{(\omega+\mu)^2}{\Delta^2}\Big)\Big\},
{\rm for} \frac{\Delta}{|\omega+\mu|}>1\nonumber\\
&&
\end{eqnarray}  
where $E(x^2)$ is the Elliptic integral of
the second kind.
In the limit of small $|\mu|$ and $T$ relative to $\Delta$, one
can find that:
\begin{equation} 
\sigma_0\sim \frac{k_B T}{\Delta}, \hbox{\quad\rm for $\mu=0$,\quad and \quad} 
\sigma_0\sim \frac{|\mu|}{\Delta}, \hbox{\quad\rm for $T=0$}.
\end{equation}
For $\mu=0$ and $T=0$, the Drude weight is
zero, as expected from the density of states. As the temperature 
increases, spectral weight is transferred to the Drude, and likewise for 
finite $\mu$.

Next, we return to the case of the interband conductivity.
Setting $t'=0$ in (\ref{eq:inter}), the factor with the hyperbolic 
functions may be moved outside the momentum sum:
\begin{equation}
\label{eq:intert0}
\sigma_{inter}(\Omega)=
\frac{\pi e^2}{\Omega}
\frac{\sinh(\beta\Omega/2)}{\cosh(\beta\mu)+\cosh(\beta\Omega/2)}
g_{inter}(\Omega),
\end{equation}
where we have defined:
\begin{equation} 
g_{inter}(\Omega)=\sum_{\bf k}\Big(\frac{\partial\epsilon_{\bf k}^-}{\partial k_x}\Big)^2
\frac{\Delta_{\bf k}^2}{E_{\bf k}^2}\delta(\Omega-2E_{\bf k}).
\end{equation}
It is now possible to obtain an analytical 
expression for $g_{inter}(\Omega)$ in the 
continuum limit:
\begin{equation}
g_{inter}=\frac{v_F^2N(0)}{\pi}\left(\frac{2\Delta}{\Omega}\right)^2
\int_0^{\pi/2}d\theta{\rm Re}\left\{\frac{\cos^2\theta}{\sqrt{1-\left(
\frac{2\Delta}{\Omega}\right)^2\cos^2\theta}}\right\},
\end{equation}
which can also be written in terms of Elliptic functions:
\begin{eqnarray}
\label{eq:ginter}
\frac{g_{inter}(\Omega)}{v_F^2N(0)/\pi}
&=&\Big\{K\Big(\frac{4\Delta^2}
{\Omega^2}\Big)-E\Big(\frac{4\Delta^2}{\Omega^2}\Big)\Big\},
{\rm for}  \frac{2\Delta}{\Omega}<1      \nonumber\\
&=&
\frac{2\Delta}{\Omega}\Big\{K\Big(\frac{\Omega^2}{4\Delta^2}\Big)
-E\Big(\frac{\Omega^2}{4\Delta^2}\Big)\Big\},
{\rm for} \frac{2\Delta}{\Omega}>1.\nonumber\\ &&
\end{eqnarray}
For small $\Omega$, $g_{inter}(\Omega)=N(0)v_F^2\frac{\Omega}{8\Delta}$
and for $\Omega \to \infty$, it is equal to 
$N(0)v_F^2(\frac{\Delta}{\Omega})^2$. Thus the interband contribution 
to the conductivity varies as 
$\beta\Omega$ at low frequency, with inclusion of the thermal factors,
 and as $\Omega^{-3}$ at large $\Omega$, 
with a singularity expected at $\Omega=2\Delta$.

The function 
$\frac{\sinh(\beta\Omega/2)}{\cosh(\beta\mu)+\cosh(\beta\Omega/2)}$, which 
multiplies $g_{inter}(\Omega)$
in Eq.~(\ref{eq:intert0}), is a universal function that shows 
the dependence of the interband conductivity  on temperature and 
chemical potential.
Taking the limit of zero temperature, 
it becomes a step function $\theta(\Omega -2|\mu|)$,{\it i.e.}, 
zero, if $\Omega<2|\mu|$ 
and one, otherwise. At finite temperature, this step function becomes smeared.
The fact that the interband conductivity starts 
abruptly at $2|\mu|$ is an important feature of the DDW model\cite{nayak}.
This feature does not arise, 
for example, in the case of
 the d-wave superconductor order parameter\cite{schachi}, rather it is 
peculiar to the DDW model
 because the unconventional density wave gap opens up at 
the antiferromagnetic Brillouin zone boundary, 
while the d-wave superconductor order parameter 
opens up at the Fermi level.

\begin{figure}[htbp]
\includegraphics[clip,width=0.45\textwidth]{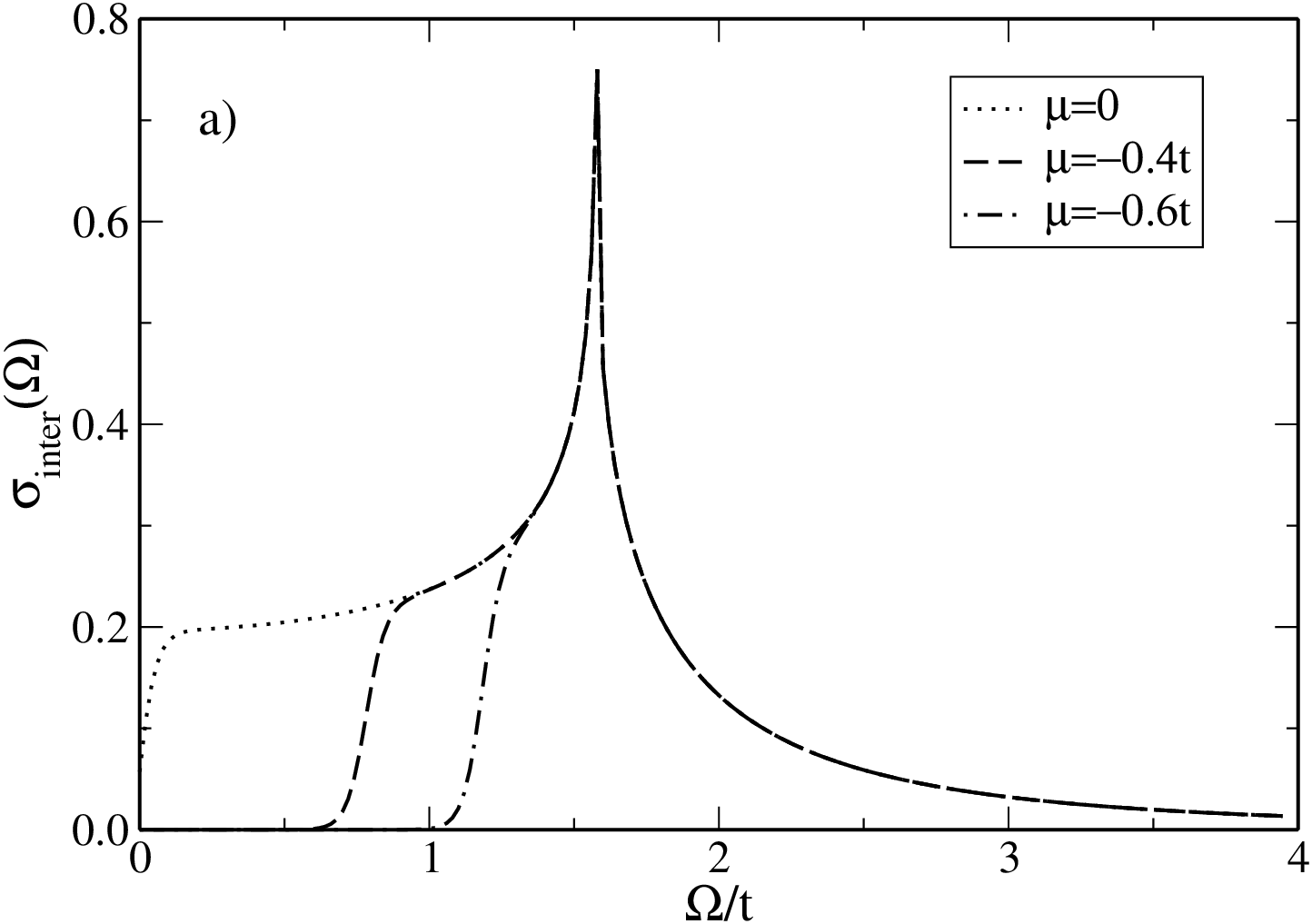}
\includegraphics[clip,width=0.45\textwidth]{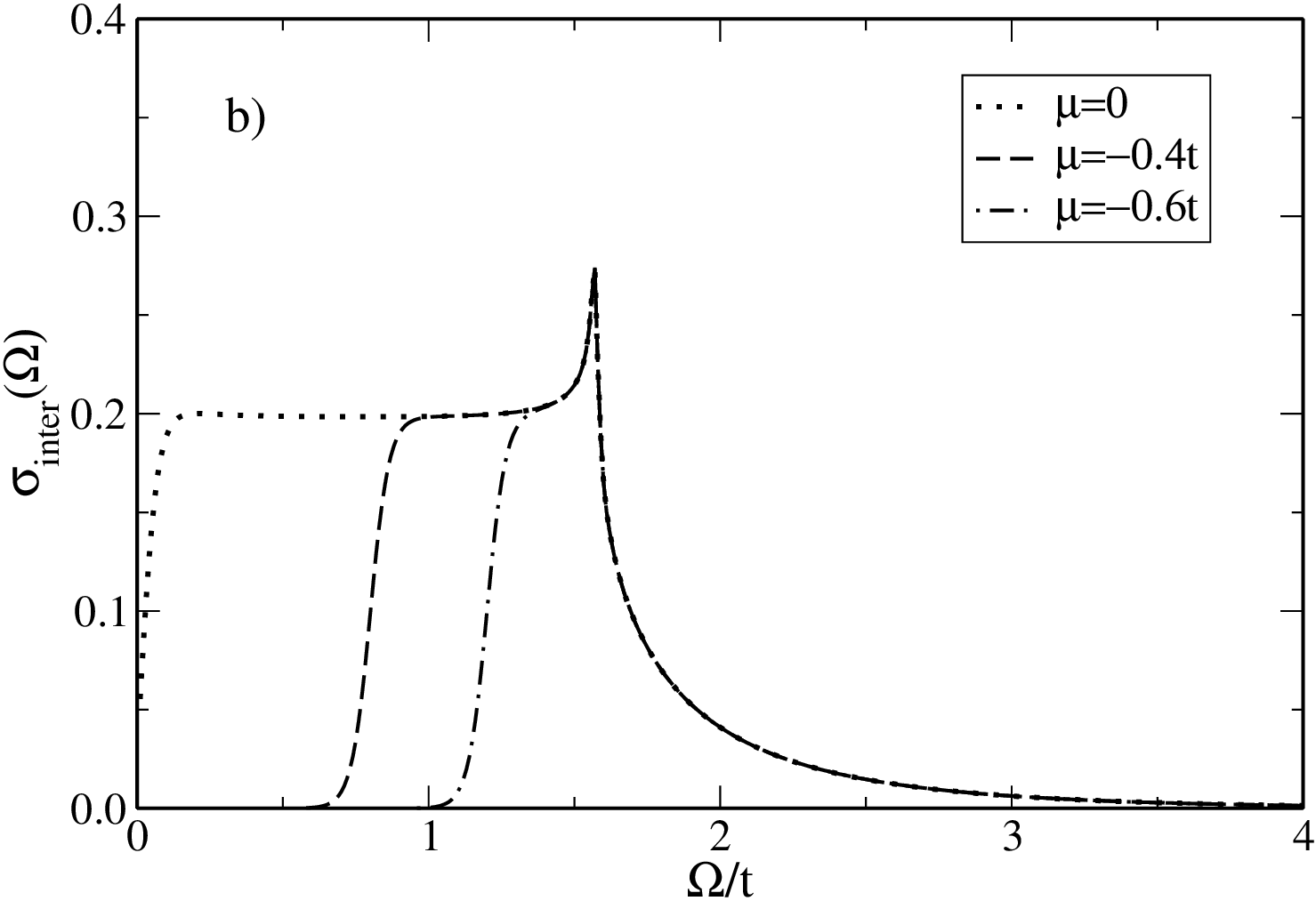}
\caption{Interband conductivity for $t'=0$, $\Delta=0.8t$ and 
$T=0.016t$ for (a) the continuum limit and (b) summing over all the  
Brillouin zone. The interband 
conductivity is shown for different values of the chemical potential $\mu$.  
In both figures, it can be seen that the interband conductivity rises 
abruptly at $2|\mu|$, and at $2\Delta$ there is a singularity 
after which it drops rapidly. In the discrete case the 
singularity is reduced.
}
\label{fig:intermu}
\end{figure}

The interband conductivity for different values of 
the chemical potential is shown in Fig.~\ref{fig:intermu}
for a fixed value of the pseudogap, $\Delta=0.8t$, and 
temperature, $T=0.016t$. Note that all figures for the conductivity
are in arbitrary units.
Fig.~\ref{fig:intermu}a has been calculated using 
the continuum limit 
given by Eq.~(\ref{eq:intert0}) and Eq.~(\ref{eq:ginter}), 
and Fig.~\ref{fig:intermu}b using Eq.~(\ref{eq:inter}) with a lattice 
size $900\times900$. The main difference 
between the continuum limit and the discrete 
sum is that the singularity around $2\Delta$ is reduced for
 the calculation performed by summing over the full Brillouin zone.
Because $\sigma_{inter}(\Omega)$ varies 
as $\beta \Omega$ when $\Omega \to 0$,
 the slope 
is set by $1/T$, which becomes very steep when $T$ is small, as 
seen in Fig.~\ref{fig:intermu} (dotted curve). On the other hand 
when $\mu$ is finite and $T\to 0$, as we have already noted, 
the thermal factor in Eq.~(\ref{eq:intert0}) provides a lower cutoff 
at $2|\mu|$, which is smeared in Fig.~3 due to finite temperature.
\vspace{0.8cm}

\begin{figure}[htbp]
\resizebox{8cm}{!}{\includegraphics{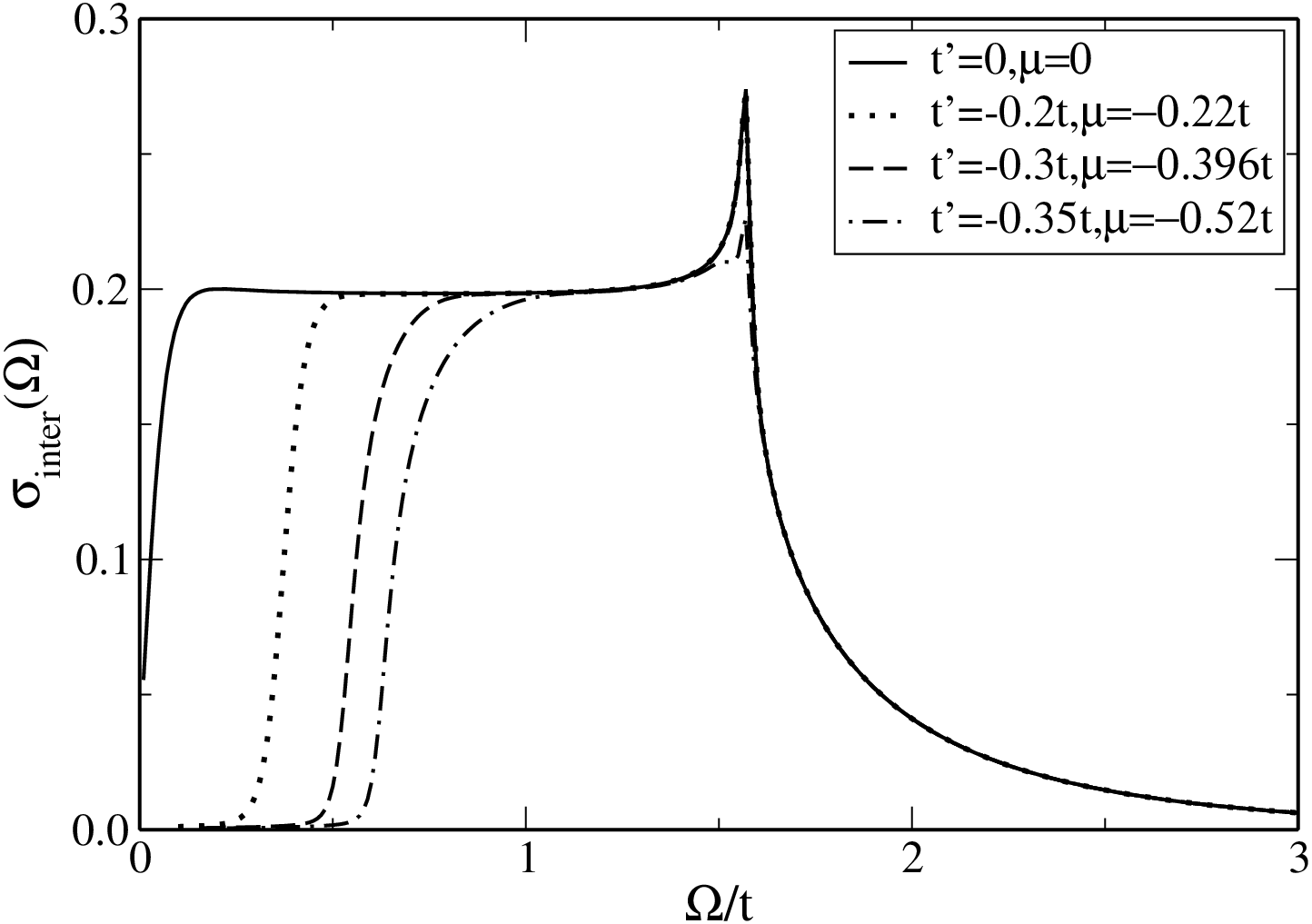}}
\caption{Interband conductivity for $x=0$, $\Delta=0.8t$, and $T=0.016t$, 
for different values of $t'$. With 
$t'$, the interband 
conductivity has a similar shape as for the case with $t'=0$, but
the lower cut off is shifted downwards in frequency.
}
\label{fig:evtintra}
\end{figure}

Next, considering the effect of $t'$ on the interband conductivity, we return
 to Eq.~(\ref{eq:inter}). 
The expression in Eq.~(\ref{eq:inter}) depends on $t'$ only through the
hyperbolic cosine where the imperfect nesting dispersion adds directly  
onto the chemical potential. Hence, the non-nesting 
term will alter the lower cutoff on the interband conductivity.  
Indeed, numerical calculation shows that the
cutoff, previously at $2|\mu|$, is shifted 
to lower 
frequency. The
$\sigma_{inter}(\Omega)$ for different values of
$t'$ is shown in Fig.~\ref{fig:evtintra}, 
which is to be compared with frame (b) of Fig.~\ref{fig:intermu}. 
In all cases, we have taken half filling, maintaining the doping 
fixed. 
The doping $x$ was defined, for temperature $T$, by:
\begin{equation}
1-x=2\int_{-\infty}^\infty d\omega \rho(\omega)f(\omega-\mu),
\end{equation}
where $\rho(\omega)$ is the density of states of Eq.~(\ref{eq:dos}),
shown in Fig.~\ref{fig:dos}, which is normalized to one state over 
the entire Brillouin zone.
The shape of the interband region 
is not much affected by the value of $t'$, but the 
cutoff at low energy is now shifted down from $2|\mu|$, as discussed above.
The surprising fact, that 
the shape of the interband contribution to the 
conductivity is not strongly affected by $t'$, may be understood
by referring to 
Fig.~\ref{fig:bands}. On comparison of frames (a) and (b), the essential
feature is that both bands shift at any value of ${\bf k}$ by the 
same amount $\epsilon^+_{\bf k}$, without
significant change in the shape, which therefore does not affect the relative
energy for vertical transitions.
\vspace{0.8cm}

We now turn to the total spectral weight, which 
has been discussed extensively as a key to 
understanding the mechanism of superconductivity.\cite{molegraaf,bontemps,benfatto} 
Having obtained expressions for inter- and intraband conductivities, 
it is straightforward to evaluate the different contributions to the 
total spectral weight, $W_{intra}=\sigma_0$ and 
$W_{inter}=\int_{-\infty}^{\infty} d\Omega \sigma_{inter}(\Omega)$, 
and determine
how they evolve with 
the various parameters of the model. 

\begin{figure}
\includegraphics[clip,width=0.45\textwidth]{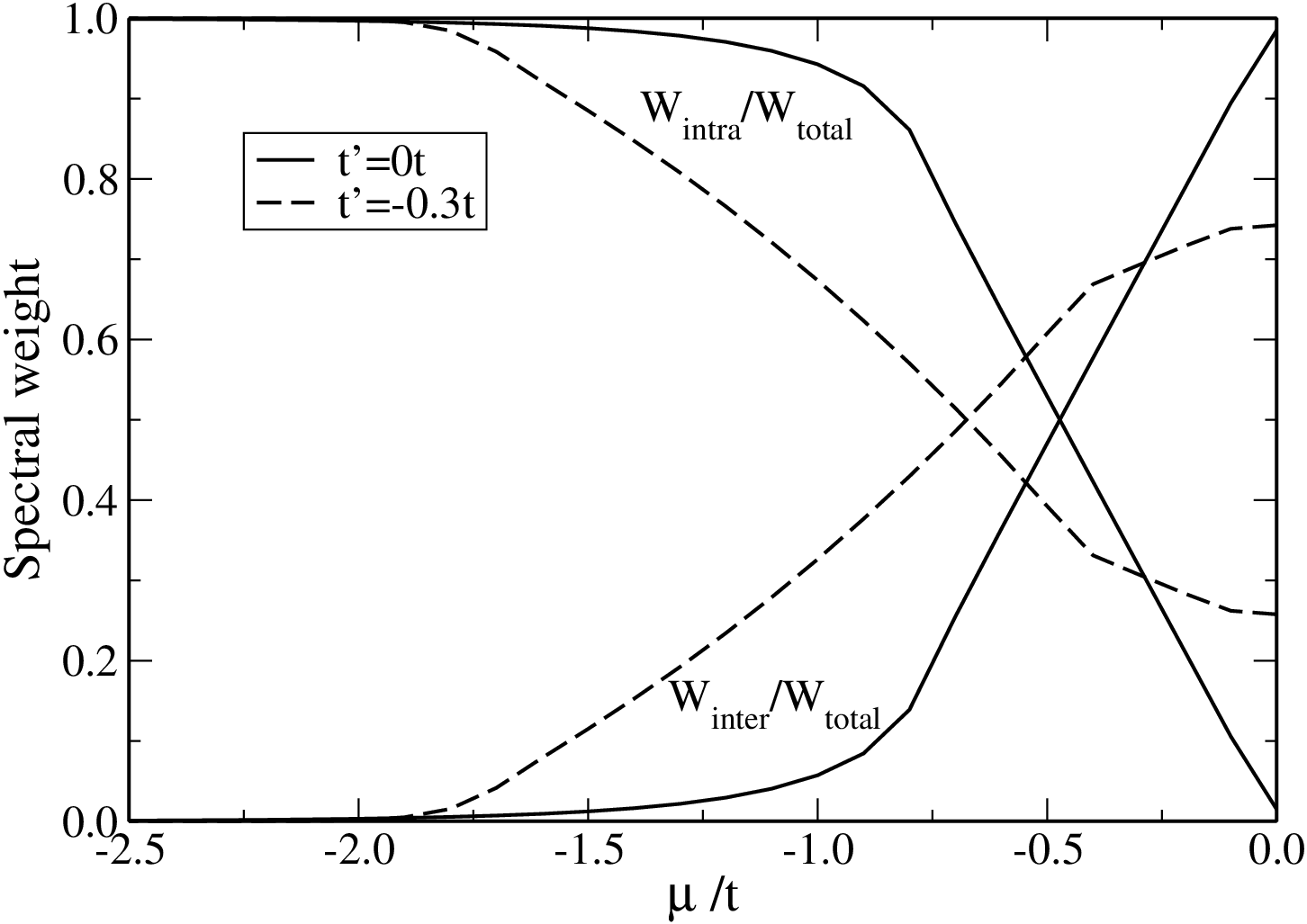}
\caption{The interband and intraband spectral weights
 normalized to the total spectral weight for $t'=0$ and $t'=-0.3t$,
for $\Delta=0.8t$.
A transfer of spectral weight from interband to intraband occurs as
the absolute value of the chemical potential increases.}
\label{fig:transfer}
\end{figure}

Integrating expression (\ref{eq:inter}) with respect to 
$\Omega$, the interband spectral weight is given by:
\begin{equation}
\label{eq:winter}
\frac{W_{inter}}{\pi e^2}=\sum_{\bf k} \biggl(\frac{\partial\epsilon_{\bf k}^-}
{\partial k_x}\biggr)^2 
 \frac{\Delta_{\bf k}^2}{E_{\bf k}^3}
\Big[f(E_{\bf k}^--\mu)-f(E_{\bf k}^+-\mu)\Big].
\end{equation}
Thus, adding Eq.~(\ref{eq:winter}) to Eq.~(\ref{eq:intra}) 
(the intraband spectral weight), we 
obtain for total spectral weight
$W_{total}= W_{inter}+\sigma_0$.
In  Fig.~\ref{fig:transfer}, it is shown that the missing spectral weight, 
in the energy range from $0<\Omega<2|\mu|$ in the
interband conductivity, is transferred to the Drude
weight as $|\mu|$ increases.
The solid curve applies to the case $t'=0$, which corresponds to the dashed 
line in Fig.~\ref{fig:dos} for the electronic density of states. 
For $\mu=0$, all the optical spectral weight is in the 
interband contribution since $\rho(\omega)=0$ at $\omega=0$, in this 
case, and there can be no intraband conductivity as a result. 
As $|\mu|$ is 
increased the intraband part increases rapidly and nearly linearly 
as does $\rho(\omega)$ versus $\omega$ of Fig.~\ref{fig:dos} in this 
range. As $|\mu|$ approaches $\Delta$ in value the growth in $W_{intra}$
becomes much less rapid and eventually saturates to being the total spectral
weight.
Thus, there is a \emph{transfer of spectral weight
from the interband region to intraband when the doping is increased}.
While the results in Fig.~\ref{fig:transfer} are based on the 
lattice sum expressions for $\sigma_{intra}$ and $\sigma_{inter}$, the 
results just described can be understood simply from Eq.~(\ref{eq:sigma0})
in the continuum limit. At $T \to 0$, 
$-\frac{\partial f(\omega) }{\partial \omega}$ becomes a delta function 
and the integral in Eq.~(\ref{eq:sigma0}) becomes equal to 
$E(\frac{\Delta^2}{\mu^2})$. This Elliptic function which describes 
the dependence of the intraband spectral weight on $\mu$ 
agrees remarkably well with the corresponding solid curve of 
Fig.~\ref{fig:transfer}. Next we consider the case for 
$t'\ne 0$. The dashed curves in Fig.~\ref{fig:transfer} apply 
to the case $t'=-0.3t$, which corresponds to the solid curve 
in Fig.~\ref{fig:dos} where we see that there is now a finite 
density of states even at $\mu=0$ and hence the intraband contribution 
starts from a finite value of approximately $0.3$ of the total weight. 
Because of the particular band structure involved, the intraband contribution 
first remains relatively flat until
 about $\mu=-0.4t$, which is the value of the chemical 
potential at half filling for the particular parameters used here,
$\Delta=0.8t$ and $t'=-0.3t$. Beyond 
$\mu \sim -0.4t$, there is a linear behaviour similar to the one 
found for $t'=0$, although with a different slope, which finally saturates 
to where all the spectral weight is intraband.  

\begin{figure}[htbp]
\includegraphics[clip,width=0.45\textwidth]{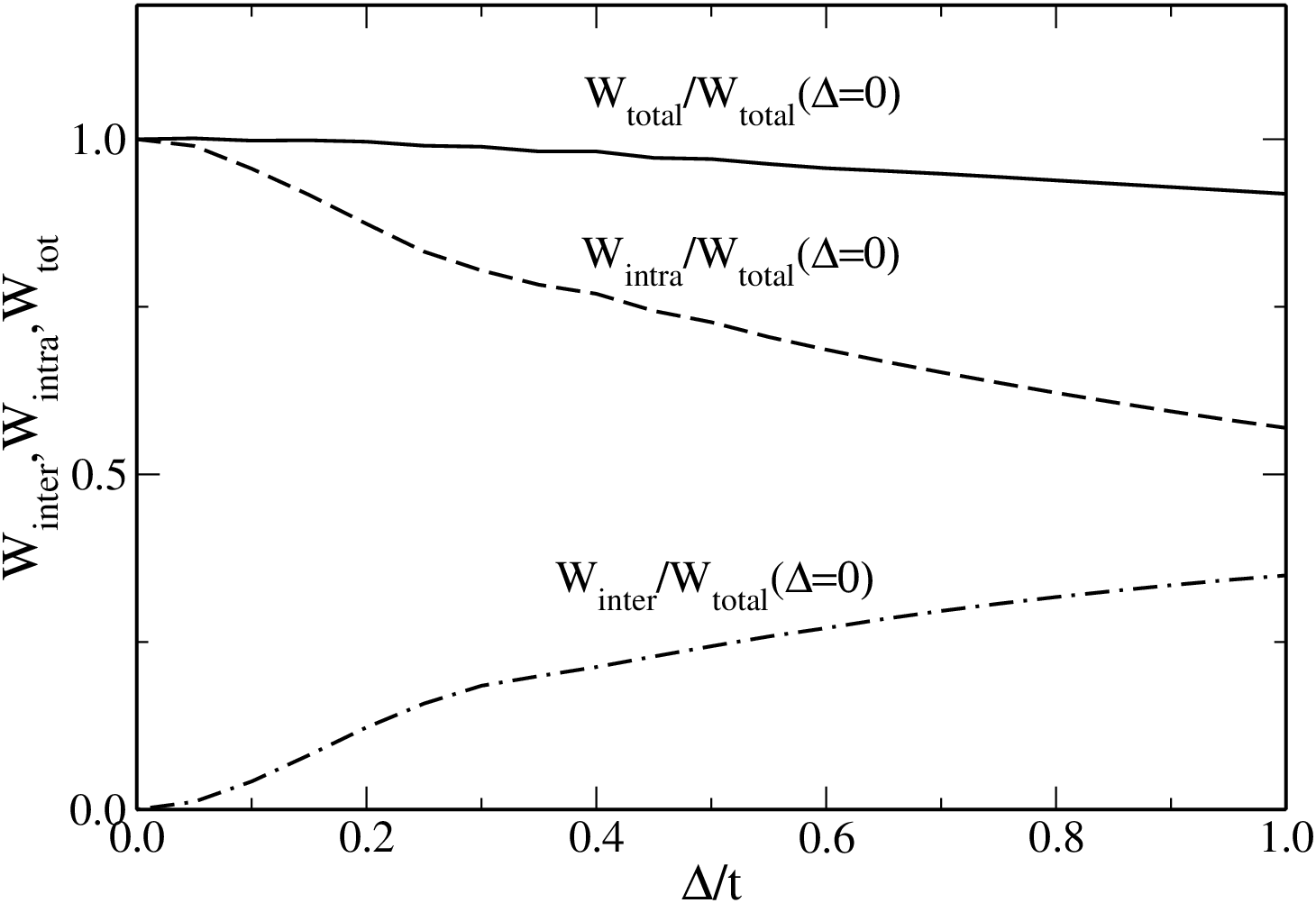}
\caption[]{Total, interband and intraband spectral weights for
fixed doping $x=0.1$. The 
total spectral weight is not changed much by the opening of
the DDW gap. 
}
\label{fig:Wvsdelta}
\end{figure}

In Fig.~\ref{fig:Wvsdelta}, we show results for $W_{inter}$ (dashed), 
$W_{intra}$ (dot-dashed) and $W_{total}$ (solid) as a function of the DDW gap
$\Delta$ in units of $t$. In these calculations, $t'=0$ and the doping was
 kept 
fixed at $x=0.1$. Remarkably, the total spectral weight
$W_{total}$ is only decreased from its $\Delta=0$ value 
by about 6\% at $\Delta=0.8t$.
As $\Delta$ increases, the amount in the intraband
component (which contains 
all the weight for $\Delta=0$) gradually decreases, however, it is still about
twice 
the amount of that 
in the interband at the largest value of $\Delta$ considered.

To conclude, we wish to emphasize that 
our most important results for the total spectral weight are: 
1) there is a transfer in the spectral weight from 
the interband conductivity at low frequencies, $0<\Omega<2|\mu|$, to 
the Drude weight at $\Omega=0$, and 2) even though a pseudogap
has opened, 
the total spectral weight is changed only by a few percent at fixed doping in 
the cases we have studied.

Other aspects of the total spectral weight were considered 
recently by Benfatto
{\it et al.}\cite{benfatto} in relation to the temperature
dependence of the total spectral weight and its relevance
to the experimental results 
by Molegraaf et al.\cite{molegraaf}, who found a very small decrease of 
$W_{total}$ with increasing temperature. Benfatto {\it et al.}'s work
also examined the effect of different choice of current operators.
Aristov and Zeyher\cite{aristov} have also briefly commented on the
spectral weight issue with respect to temperature variation.
A feature of a recent work of Benfatto {\it et al.}\cite{benfatto2}
is the examination of the role of ward identities and
vertex corrections on the temperature
dependence of the optical sum rule, where
they find that the total spectral weight increases with decreasing temperature.
While there is some difference between their total spectral weight and ours,
upon comparison with the frequency-dependent optical conductivity,
we find very little difference between our calculation and the one shown
in their paper. They are virtually the same, with only a small quantitative 
difference at high frequency, and all the features discussed here
remain intact. This indicates that these corrections, 
while important for the total spectral weight, 
may be neglected, as is common, 
for calculations of the
frequency-dependent optical conductivity. Their work, done concurrently, is
complementary to ours in that they examine primarily the issue of total
spectral weight and we emphasize the details of
frequency-dependent conductivity.
Our aim in this work is to determine how the
total spectral weight is distributed in the optical conductivity
and where it transfers to upon the opening of the pseudogap. 

\section{Electronic Raman scattering}
\label{sec:raman}
Now that we understand the crucial role of the chemical potential 
in the optical conductivity (indicating the frequency above which
the interband response is different from zero),
we examine its effects on the Raman response. The Raman response has 
the advantage that different regions of the Brillouin zone 
may be selected via  
the choices of incoming and outgoing photon polarizations. In contrast,  
the optical conductivity is an average over all the Fermi surface.
Hence, Raman experiments are relevant for unconventional density 
waves since they can provide information on the symmetry of the 
order parameter. 

In the point group $D_{4h}$ of the square lattice, 
we have three different channels:$B_{1g}$, $B_{2g}$, and $A_{1g}$. 
The $B_{1g}$ channel projects out the antinodal region of the 
Fermi surface, the $B_{2g}$, the nodal region, and the $A_{1g}$ is a 
weighted average over the entire Brillouin zone. In evaluating the
Raman response for these different channels, we will discuss 
the continuum limit for simplicity.

The real part of the optical conductivity multiplied by frequency,
{\it i.e.} $\Omega{\rm Re}\{\sigma(\Omega)\}$, is
related to the imaginary part of the Raman susceptibility\cite{shastry},
$\chi_{\gamma_\nu}$, in the case where the $v_F^2$ in the conductivity
is replaced by $\gamma_\nu^2$, where 
$\gamma_\nu$ is the Raman vertex for a particular channel.
In the absence of impurities, the Drude part of the 
Raman signal is zero and the interband part turns out to be:
\begin{equation}
\label{eq:raman}
{\rm Im}\{\chi_{\gamma_\nu}(\Omega)\}=
\frac{\sinh(\beta\Omega/2)}{\cosh(\beta\mu)+\cosh(\beta\Omega/2)}
g^{\gamma_\nu}_{inter}(\Omega),
\nonumber\\
\end{equation}
where $\nu$ stands for the three different channels mentioned above:
$A_{1g}$, $B_{1g}$ and $B_{2g}$, and we
have defined:
\begin{eqnarray} 
\label{eq:graman}
g^{\gamma_\nu}_{inter}(\Omega) =
\frac{N(0)}{2\pi}\left(\frac{2\Delta}{\Omega}\right)^2\int_0^{\pi/2} 
d\theta 
\gamma_\nu(\theta/2)^2\nonumber\\
\times{\rm Re}\left\{\frac{\cos^2\theta}{\sqrt{1-\left(
\frac{2\Delta}{\Omega}\right)^2\cos^2\theta}}\right\}
\end{eqnarray}
The expressions for the unrenormalized Raman vertices in the continuum 
limit are:
\begin{eqnarray}
\gamma_{A_{1g}}({\theta})&=&a \cos(4\theta), \\
\gamma_{B_{1g}}({\theta})&=&b_1 \cos(2\theta), \\
\gamma_{B_{2g}}({\theta})&=&b_2 \sin(2\theta),
\end{eqnarray}
\begin{figure}[htbp]
\includegraphics[clip,width=0.45\textwidth]{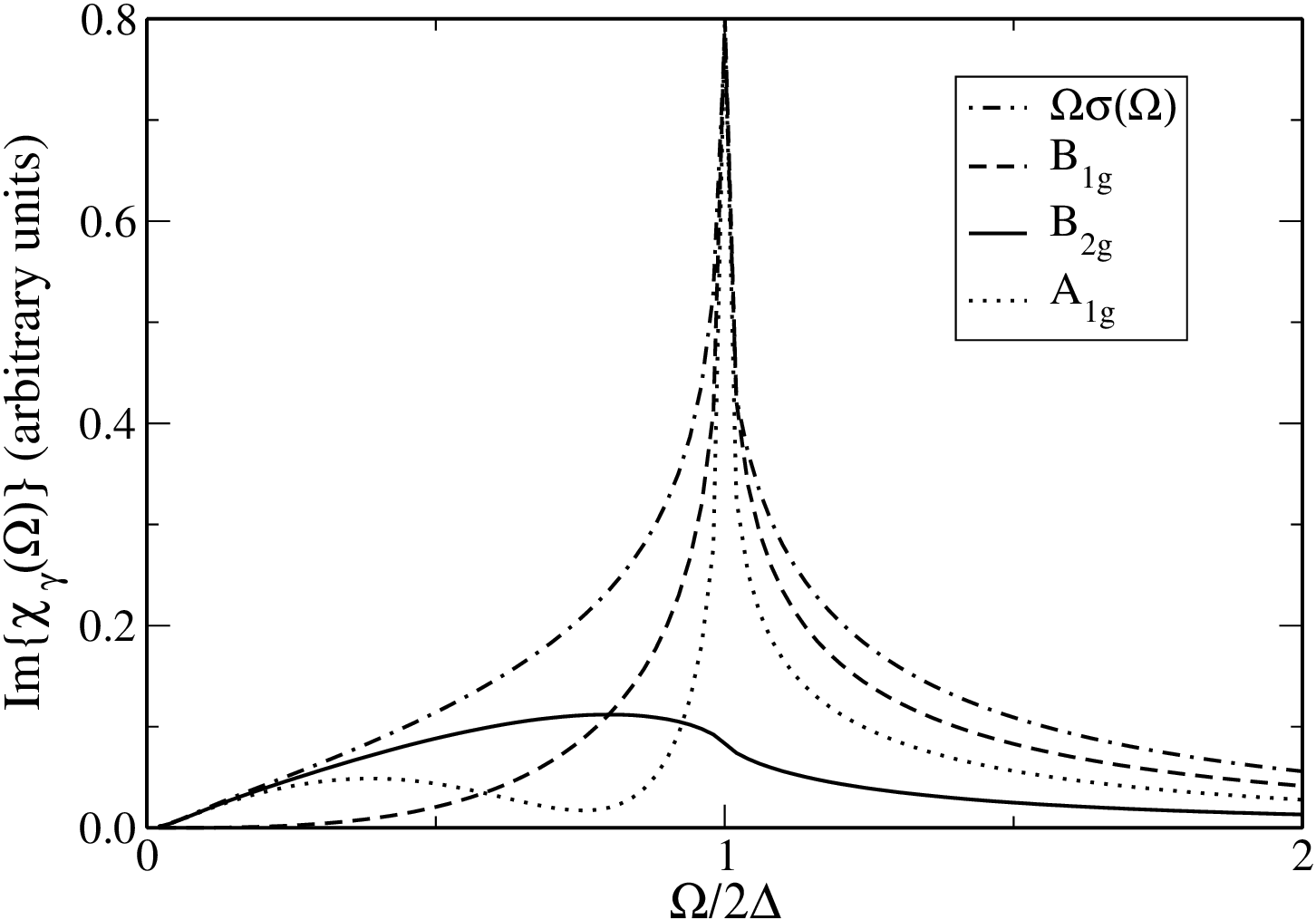}
\caption{Raman signal for $A_{1g}$ (dotted), $B_{1g}$ (dashed), 
and $B_{2g}$ (solid) channels
for $\Delta=0.5t$, $\mu=0$ and $T=0.016t$.
At $2\Delta$, there is a singularity in both $A_{1g}$ and $B_{1g}$ channels,
and a small kink in $B_{2g}$.
$\Omega{\rm Re}\{\sigma(\Omega)\}$ (dash-dotted) is also shown for comparison. 
All the signals are on the same scale.
}
\label{fig:ramanmu0}
\end{figure}
\vspace{0.8cm}
and we have absorbed additional constant factors in the $a$ and 
$b$'s which are arbitrary.
The integrals in Eq.~(\ref{eq:graman}) can be written in terms of 
Elliptic integrals which have the same form as for a d-wave 
superconductor.\cite{dwayne}
The result is shown in Fig.~\ref{fig:ramanmu0} for $\mu=0$ and in 
Fig.~\ref{fig:ramanmu} for $\mu=-0.25t$.
To better illustrate the role of the chemical potential in providing  
a lower frequency cutoff on the 
Raman signal, we 
have used $a=b_1=b_2=1$. This is certainly not realistic.
In fact, in the tight-binding approximation, 
the signal in $B_{2g}$ is proportional to $t'$ 
instead of $t$,\cite{t} and as a consequence we expect that 
the $B_{2g}$ signal 
will be much smaller than 
the other ones. In the low 
frequency regime, the $A_{1g}$ (dotted) and the  $B_{2g}$ (solid) signals 
are proportional to $\omega/\Delta$, while $B_{1g}$ (dashed) is proportional to 
$(\omega/\Delta)^3$. For comparison we have also included 
our results for $\Omega{\rm Re}\{\sigma(\Omega)\}$ (dash-dotted curve) based 
on Fig.~\ref{fig:intermu}a.
\begin{figure}[htbp]
\includegraphics[clip,width=0.45\textwidth]{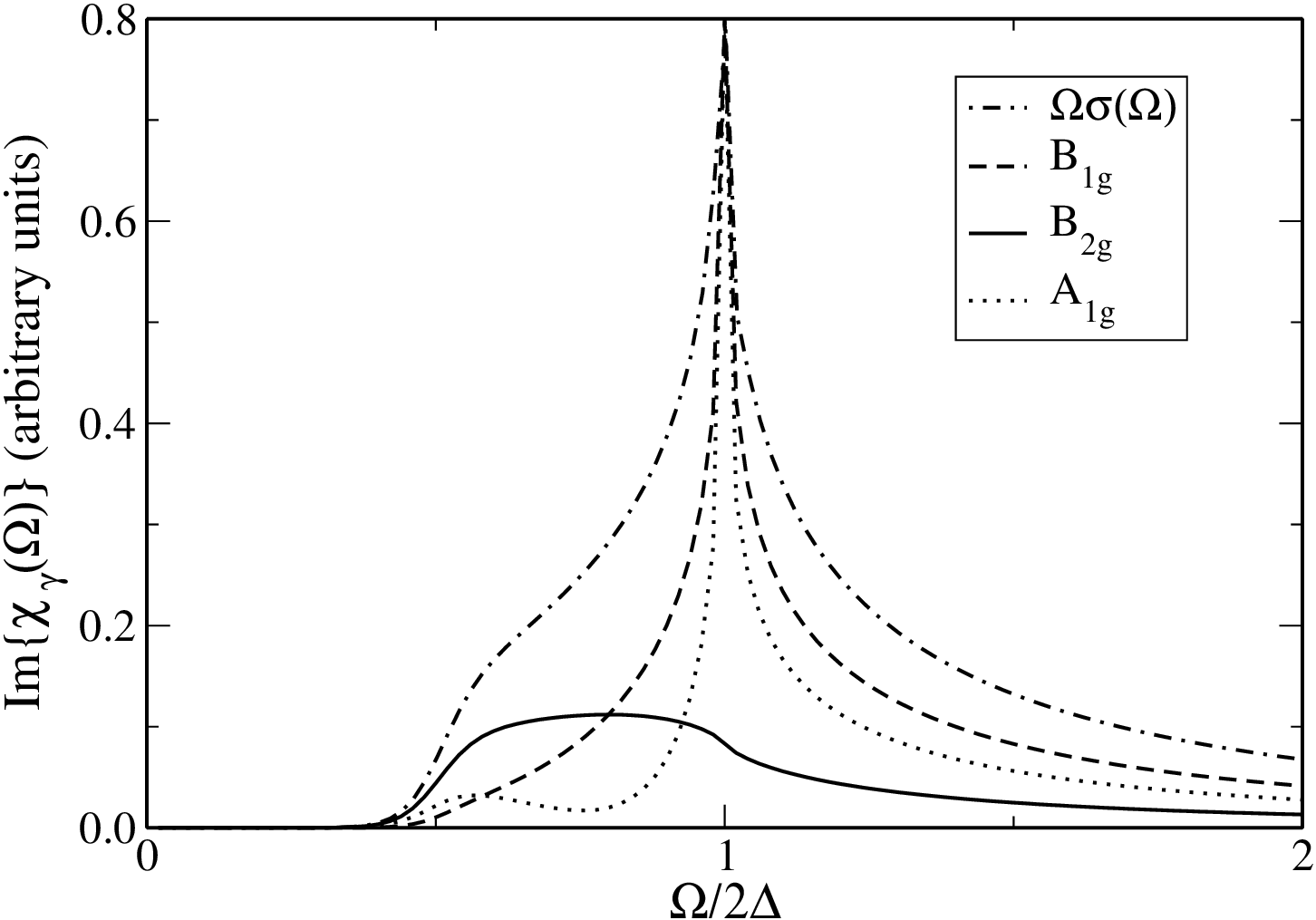}
\caption{Raman signal for the $A_{1g}$, $B_{1g}$, and $B_{2g}$ channels
for $\Delta=0.5t$, $\mu=-0.25t$, and $T=0.016t$. 
The chemical potential cutoff is 
easily seen in $B_{2g}$ 
and $A_{1g}$. The magnitude of the signal has been set to be the same 
in all cases, although $B_{2g}$ will be, in reality, 
smaller. $\Omega{\rm Re}\{\sigma(\Omega)\}$ is also shown for comparison.
}
\label{fig:ramanmu}
\end{figure}
The Raman signal has recently been studied by 
Zeyher and Greco\cite{greco} within the $t-J$ model, 
with particular attention paid to the renormalization of the Raman vertices
and to the role of superconductivity.
Here, we are only interested in the mean field prediction for 
the region of the phase diagram where the DDW order prevails.

From Fig.~\ref{fig:ramanmu} we conclude that the effect of the $2|\mu|$
cutoff 
is easily seen in the $A_{1g}$ and the $B_{2g}$ signals, but less so for 
the $B_{1g}$ channel,
 which is already quite small in magnitude at small $\omega$.
This is because the $B_{1g}$ symmetry probes the 
antinodal quasiparticles where the pseudogap is largest.
\section{Scattering}
\label{sec:scatt}
We now turn to the problem of scattering by impurities and inelastic 
scattering. We study only the optical conductivity since the results 
can be easily extrapolated to the Raman response.
To include quasiparticle scattering, we replace the delta functions inside 
the spectral functions $A_{ij}({\bf k},\omega)$ in 
Eq.~(\ref{eq:A}), with 
Lorentzians of the form:
\begin{equation}
\Gamma({\bf k},\omega)^\pm=\frac{1}{\pi}\frac{\gamma(\omega)}{(\omega+\mu-E_{\bf k}^\pm-
\rm Re\Sigma({\bf k},\omega))^2+\gamma(\omega)^2},
\end{equation} 
where $\gamma(\omega)=-\rm Im\Sigma({\bf k},\omega)$ and 
$\Sigma({\bf k},\omega)$ is the self-energy due to impurities 
or coupling to inelastic scattering.   
These $A_{ij}$ are inserted in the optical conductivity 
given by Eq.~(\ref{eq:resigma}).

We study two cases, 
the scattering rate equal to a constant: 
$\gamma(\omega)=\eta$, and the scattering 
rate of the marginal Fermi liquid:
$\gamma(\omega)=\eta+\pi\lambda|\omega|$ where $\eta$ and $\lambda$ 
are constants. 

\subsection{$\gamma(\omega)=\eta$}
If $\gamma(\omega)=\eta$, then $\rm Re\Sigma({\bf k},\omega)=0$. 
The effect of 
adding a constant scattering rate is that the interband contribution 
can no longer be separated from the Drude contribution. Still, these 
two contributions can 
be clearly discerned in the real part of the conductivity 
for a large range of parameters. 
\begin{figure}[htbp]
\includegraphics[clip,width=0.45\textwidth]{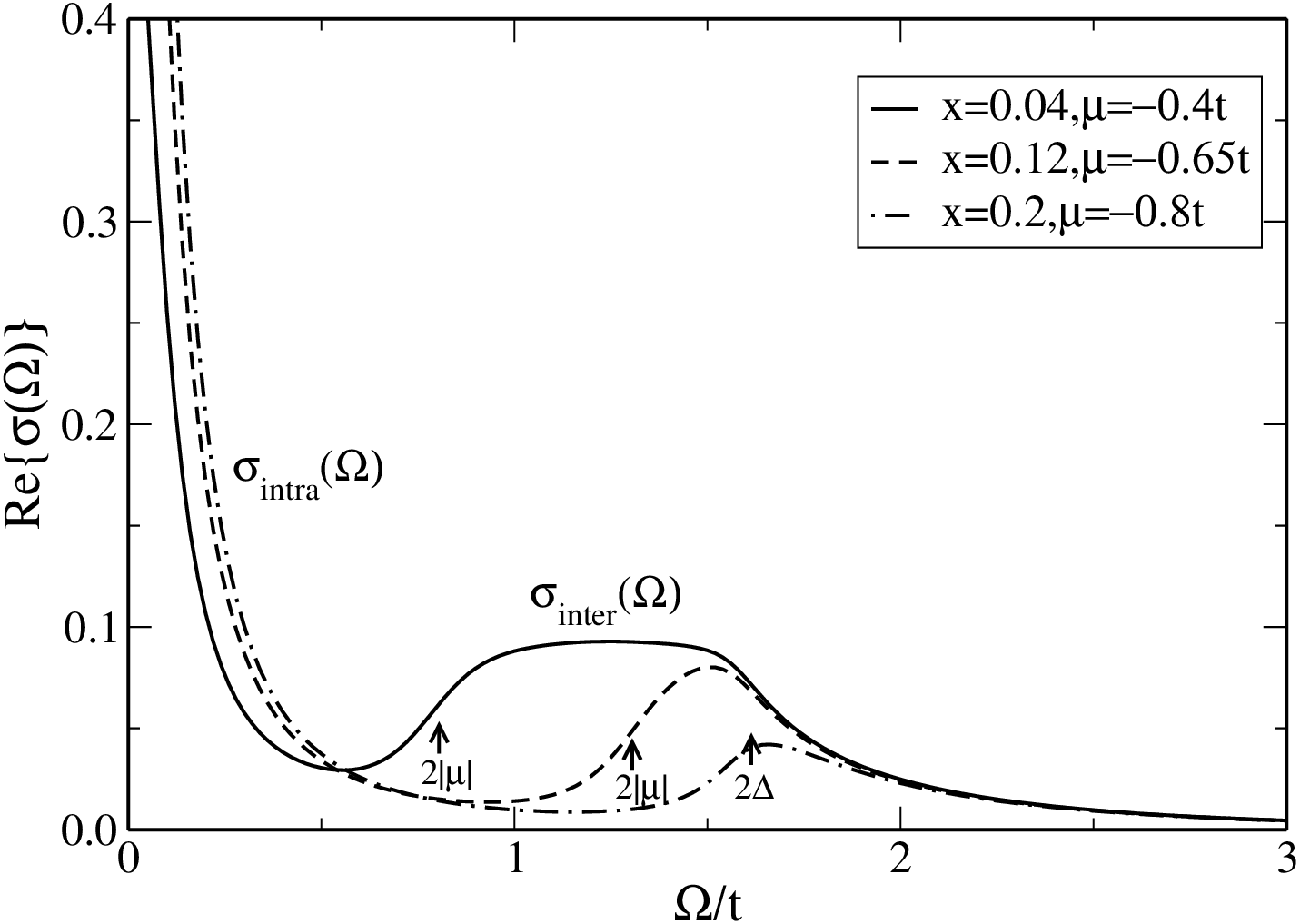}
\caption{Evolution of the real part of the
optical conductivity with doping for
a fixed gap, $\Delta=0.8t$, $t'=0$, constant scattering
rate $\eta=0.05t$, and $T=0.016t$. The values $2\Delta$
and $2|\mu|$ are indicated by arrows. As the doping
is increased, the lower edge corresponding to $2|\mu|$
is moving to higher frequency and the
Drude weight is increasing.
}
\label{fig:sigmaevx}
\end{figure}
For the numerical calculations of the 
optical conductivity given by Eq.~(\ref{eq:resigma}), we have used a lattice
size of 300x300 for the Brillouin zone sum.
The evolution with doping of the 
$\rm Re\{\sigma(\Omega)\}$ is shown in
Fig.~\ref{fig:sigmaevx}. 
In this figure, we can clearly see
the redistribution of the spectral weight 
between the Drude contribution and the interband contribution when 
the doping is varied. In all these curves, solid for $x=0.04$, 
dashed for $x=0.12$ and dash-dotted for x=0.2, the lower edge 
of the interband region is at $2|\mu|$ as indicated by arrows with 
some smearing due to temperature and impurity scattering. This 
edge moves upwards with increasing doping. There is little 
change however in the upper edge of the band at approximately 
$2\Delta$. For the dash-dotted curve, the interband 
region is almost completely gone and has been transfered to 
the Drude peak which has increased optical strength as compared 
to the solid curve. The effect of varying the impurity parameter
$\eta$ is very straighforward and, therefore, not shown. As $\eta$ 
is increased the Drude component at low frequency is broadened in
width and the peak drops, likewise the interband contribution is further
smeared out and reduced in height.

\begin{figure}
\includegraphics[clip,width=0.45\textwidth]{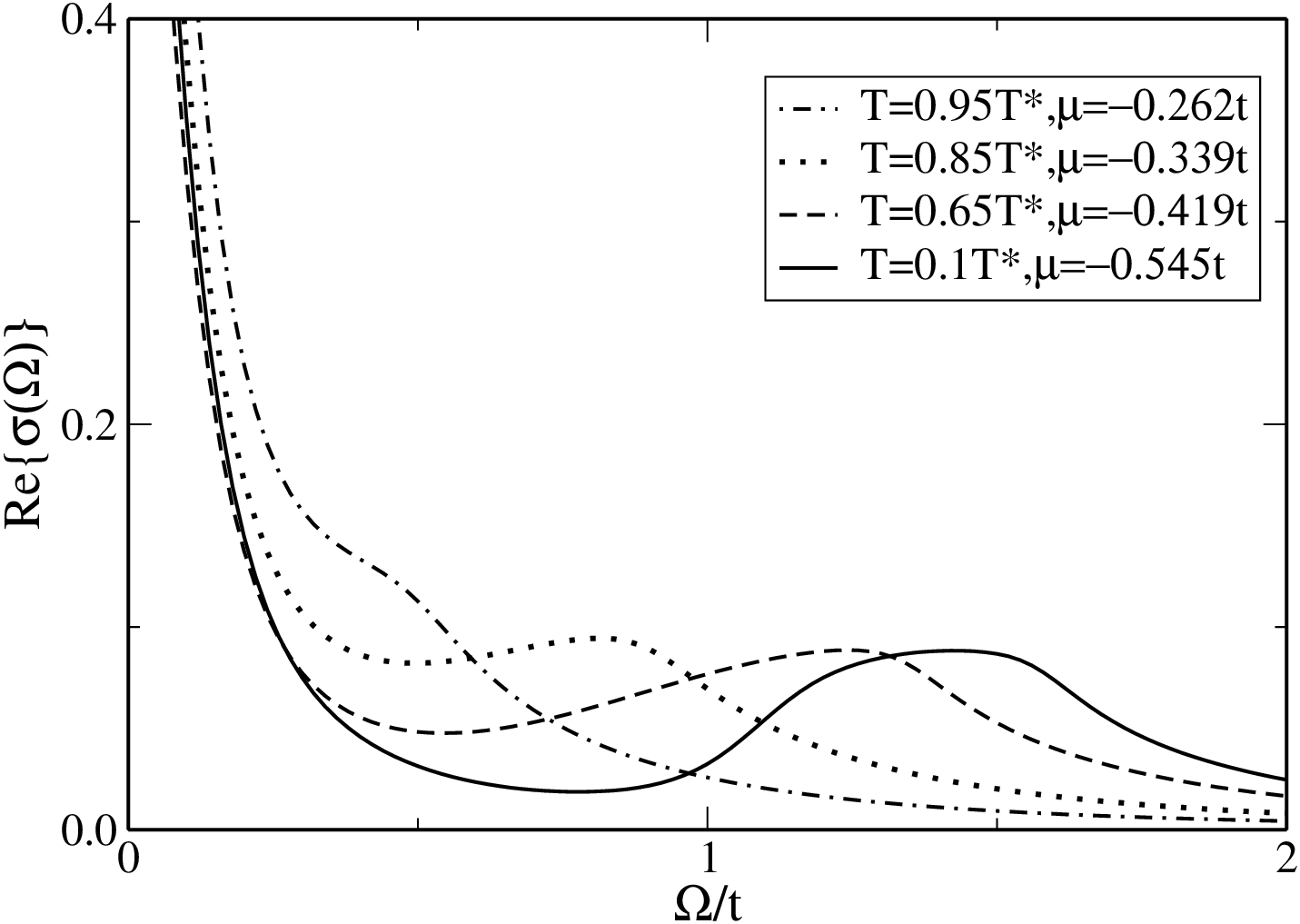}
\caption{Evolution of the real part of the optical conductivity with
temperature for $\Delta=0.8t$, $t'=0$, $\eta=0.05t$ and $x=0.08$. The edge at
$2\Delta$ is moving toward lower frequency with 
increasing temperature because of the temperature
dependence of $\Delta$. 
The Drude peak
is increasing with temperature, although the opposite behaviour is 
found when inelastic scattering is included.
}
\label{fig:sigmaevTemp}
\end{figure}

We now turn to the question of how $\rm Re\{\sigma(\Omega)\}$ varies with 
temperature for fixed doping. For this calculation we have used a fitted
dependence of the gap with temperature\cite{benfatto}:
$\Delta(T)=(1-\frac{1}{3}(T/T^*)^4)\sqrt{1-(T/T^*)^4}$, where we have 
chosen $T^*=0.16t$. 
As the temperature is 
increased, the gap seen as the upper edge of the interband transitions 
moves to lower frequencies because the 
pseudogap closes. Secondly, 
the depression between low frequencies and $2|\mu|$ 
is filled in with increasing temperature.
More importantly, the
Drude weight increases 
with increasing temperature because of the collapse of the interband 
region. The DC conductivity displays a typical semiconducting 
behaviour with resistivity decreasing with increasing temperature. 
However, in these calculations we have kept the 
scattering rate constant. In any realistic case, the inelastic scattering 
will increase with $T$. This can be modelled by a temperature 
dependent scattering rate which increases with temperature, therefore 
reducing the DC conductivity, and as a result there can be a
change of its behaviour from semiconducting 
to metallic. Such inelastic scattering would modify the result of 
Fig.~\ref{fig:sigmaevTemp}.
It would broaden the Drude peak relatively to what is seen 
and fill in the 
region between the intraband  and interband contributions even more,
obscuring the $2|\mu|$ cutoff.

Experimentally, the spectral weight distribution at low frequency 
and its temperature dependence has been studied in 
reference \cite{bontemps}. For their underdoped sample, the
spectral weight 
in the pseudogap region, does not move to 
lower frequencies with increasing temperature as is predicted in DDW
theory, but rather it has the opposite behaviour;
 the spectral weight at low frequencies moves
 to the Drude with decreasing temperature.\cite{puchkov,bontemps}
This behaviour could only be obtained in the DDW model with strong inelastic
scattering.

\begin{figure}[htbp]
\includegraphics[clip,width=0.45\textwidth]{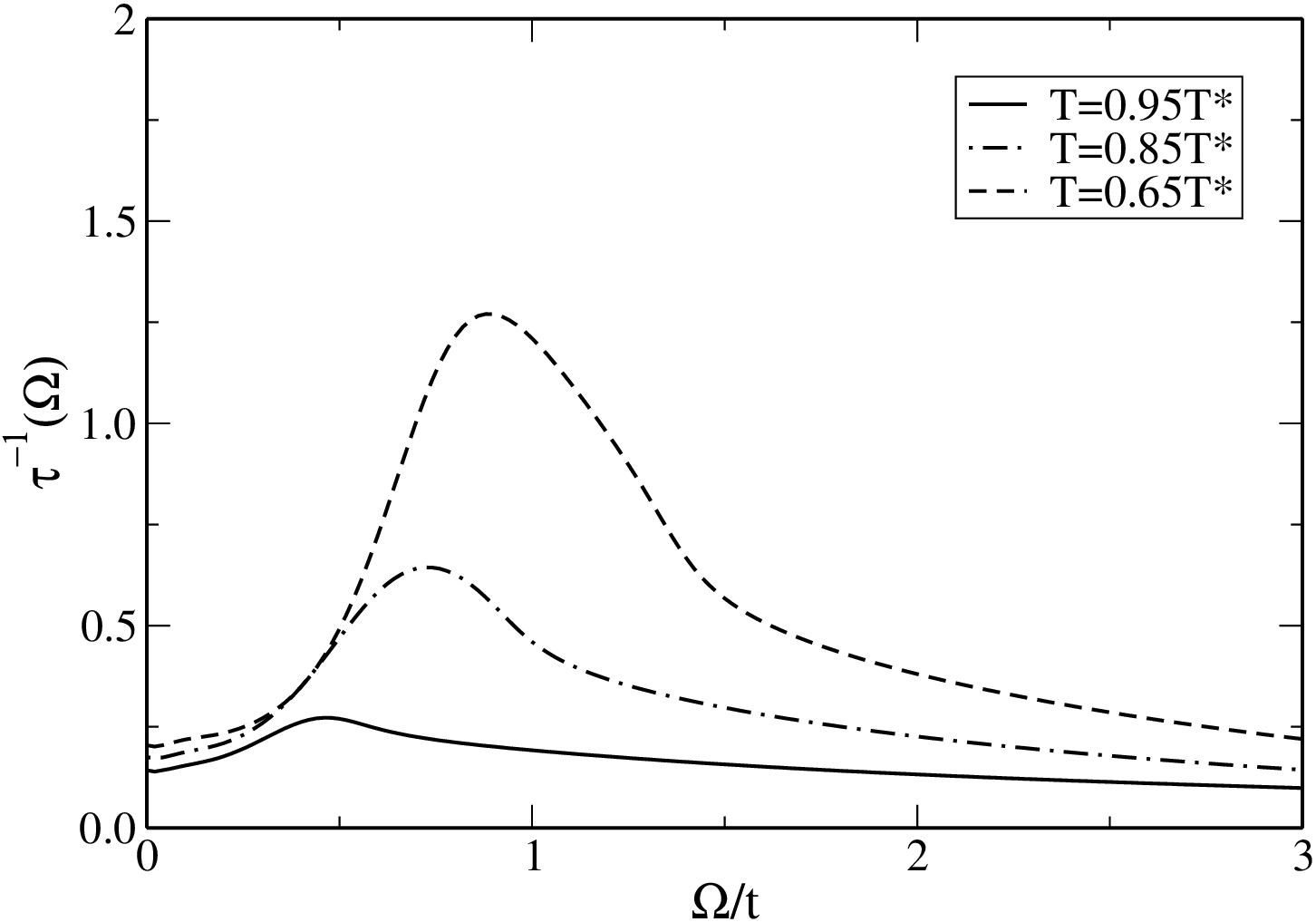}
\caption{Evolution of the optical scattering rate with
temperature for $x=0.08$, $\Delta=0.8t$, $t'=0$, and
$\eta=0.05t$. At low temperatures, there is a strong
peak caused by the interband processes,
which washes out with increasing temperature, and the
scattering rate eventually recovers 
the pure Drude behaviour. 
}
\label{fig:tau} 
\end{figure}

Next, we calculate another important quantity that is obtained 
in optical experiments, 
namely the optical scattering rate. For this, we use the
extended Drude model:\cite{tanner-timusk,puchkov}
\begin{equation}
\label{eq:taudef}
\tau^{-1}_{op}(\Omega)=\frac{\Omega_p^2}{4\pi}\frac{{\rm Re}\{\sigma(\Omega)\}}
{{\rm Re}\{\sigma(\Omega)\}^2+{\rm Im}\{\sigma(\Omega)\}^2},
\end{equation}
where we have calculated numerically
the $\rm Im\{\sigma(\Omega)\}$ by Kramers-Kronig
relations, 
and for the plasma frequency we have used that derived
from the Drude weight in the case of  
$\Delta=0$. The result is shown in Fig.~\ref{fig:tau}.
The most striking feature in this figure is a peak 
which arises from the interband 
contribution. 
This is an important result since, in principle, it can be verified 
experimentally. We note that as the temperature increases, the size 
of the peak decreases in magnitude and shifts to lower energy 
before disappearing completely at the pseudogap temperature $T^*=0.16t$.
Also the position of the peak in the scattering rate of 
Eq.~(\ref{eq:taudef}) is not at the same position 
as the peak in the real part of the conductivity. It is lower.
In particular for the solid curve in Fig.~\ref{fig:tau}, which is 
the highest temperature shown, the peak falls below 
$0.5t$ while the corresponding structure in the real part 
of the conductivity in Fig.~\ref{fig:sigmaevTemp} is nearer 
$0.8t$, although in that case it cannot be characterized 
simply as a peak. Later, we will comment on the origin of these 
shifts, between the scattering rate and the optical conductivity, of the 
interband structure. A second feature to be noted in 
Fig.~\ref{fig:tau} is that the scattering rates at 
$\omega \to 0$ are considerably larger than twice 
the value of the impurity scattering rate that we have 
used. This feature will also be addressed at the end 
of this section. Finally, as discussed before, if the impurity
parameter $\eta$ is increased, the conductivity is broadened
and the effect on the peak seen in the scattering rate is for it
to be broadened and reduced.
Recently a peak in $\tau_{op}^{-1}(\omega)$ in the region of 
$400$ to $3500$ ${\rm cm}^{-1}$ has been observed in the 
cobaltates\cite{timuskCo}.
This peak reduces and eventually vanishes with increasing temperature. 
There appears to be a corresponding peak in the conductivity, as well,
at higher frequency (400 to 6000 cm$^{-1}$), not unlike what we
find in our model.
\vspace{0.8cm} 

\subsection{Inelastic scattering}
To present a more realistic comparison with experiments, we should include the 
inelastic scattering. It is important to check whether or not the strong 
feature found above, in the optical scattering rate, will be washed 
out by the inelastic scattering. We can do this using the self-energy of the 
marginal Fermi liquid\cite{mfl} in our spectral functions 
$A_{ij}({\bf k},\omega)$.
This phenomenological model assumes 
that the excited charge carriers 
interact with a wide, flat spectrum of excitations over 
$T<\omega<\omega_c$, where $\omega_c$ is a high energy
cutoff of the spectrum.
This model was introduced to explain the observed linear 
behaviour in the scattering
rate $\Gamma \sim {\rm max}(|\omega|,T)$ 
measured in the optical conductivity, as well as in the
resistivity, and in the Raman cross-section.\cite{varma,tanner-timusk}
\vspace{0.5cm}
\begin{figure}[t]
\includegraphics[clip,width=0.45\textwidth]{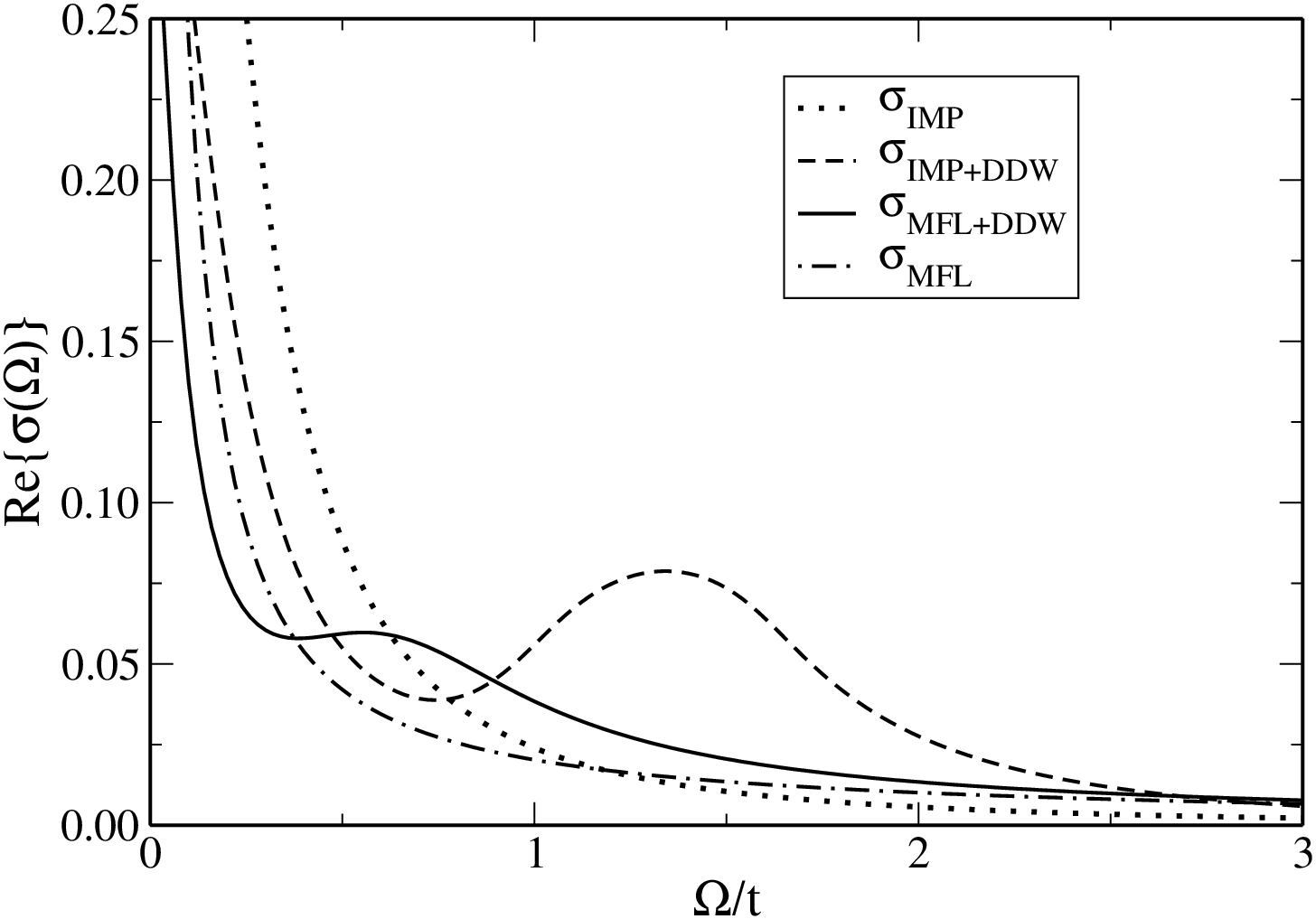}
\caption{Comparison of the real part of the optical conductivity
including the marginal Fermi liquid self-energy (dash-dotted curve) 
with the case where
the scattering rate is constant (dotted curve): $\gamma(\omega)=\eta$. Here,
$\Delta=0.8t$, $\mu=-0.5t$, $t'=0$, $\pi\lambda=0.5$, $\eta=0.1t$,
and $T=0.016t$. Two
characteristics of the marginal Fermi liquid are evident:
1. there is more spectral weight
at higher frequencies and 2. the low frequency peak is narrowed 
in comparison with the dotted curve. Also shown are the results with
a DDW included.
}
\label{fig:compmar}
\end{figure}
The marginal Fermi liquid ansatz is:\cite{tanner-timusk,mfl}
\begin{equation}
\Sigma_{MFL}(\omega)=2\lambda \omega \ln \frac{|\omega|}{\omega_c}-i\pi
\lambda |\omega|,
\label{eq:mflself}
\end{equation}
where $\lambda$ is a dimensionless coupling constant.
\begin{figure}
\includegraphics[clip,width=0.45\textwidth]{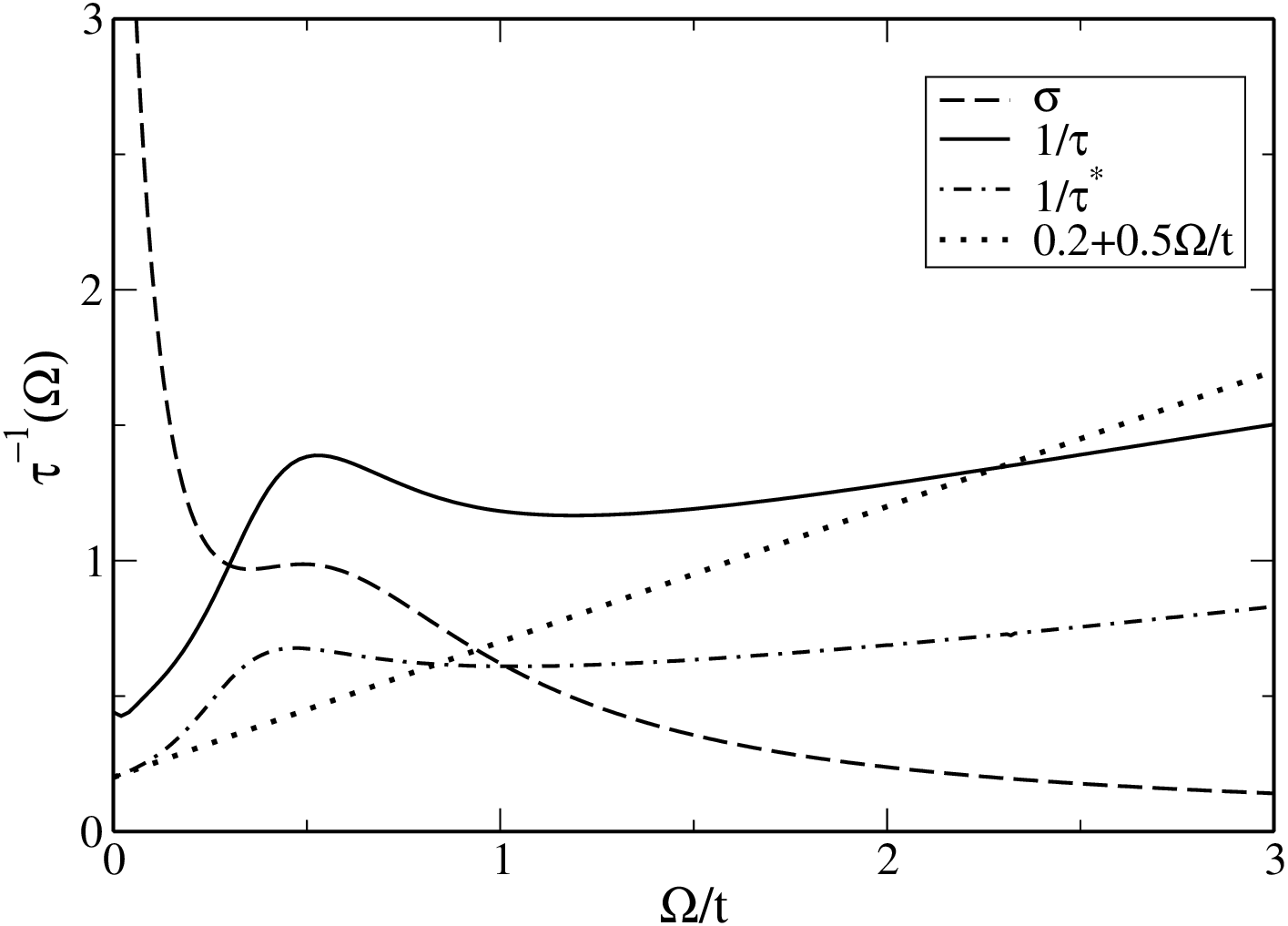}
\caption{Optical scattering rate 
for the MFL with the DDW. Shown is $1/\tau(\Omega)$ (solid line)
 for the case of $\Delta=0.8t$,
$\mu=-0.5t$, $t'=0$, $\pi\lambda=0.5$ and $\eta=0.1t$, calculated
from the conductivity curve given by the solid
line in Fig.~\ref{fig:compmar}.
The dashed line is an approximate fit to the conductivity curve
 of Fig.~\ref{fig:compmar} using the model of a MFL plus
a displaced oscillator, as discussed in the text, and from this
$1/\tau^*(\Omega)$ is 
calculated (dash-dotted line) and can be seen to 
be related to $1/\tau(\Omega)$ of Eq.~(\ref{eq:taudef}) by a multiplicative 
constant.
}
\label{fig:lorenz}
\end{figure}

Results based on the MFL self-energy are 
shown 
in Fig.~\ref{fig:compmar}, where they are also compared 
with previous results for impurities alone ({\it i.e.}, 
a constant scattering rate).
The dotted line is for $\eta=0.1t$ and the dash-dotted curve is for
$\eta=0.1t$ plus the MFL contribution to the self-energy. 
On comparing these two
curves, both calculated at temperature $T=0.016t$, we note that considerable
optical spectral weight is lost at small $\Omega$ in the dash-dotted curve,
as compared to the dotted Drude curve. 
There is a  crossing at $\Omega/t$ slightly
above 1 and for the MFL
 there is a long tail in the mid-infrared region of the optical
spectrum, extending to high $\Omega$ and staying well above the Drude 
curve. These tails are an essential feature of the conductivity in the
cuprates, an indication of a strong electron-boson inelastic contribution
(Holstein band). The depletion of the optical weight at small $\Omega$
reflects the fact that the quasiparticle spectral weight in the MFL, 
$Z(\omega)=(1-\frac{\partial {\rm Re} \Sigma}{\partial \omega})^{-1}$, 
varies as
$(2\lambda \ln \frac{\omega_c}{|\omega|})^{-1} $ at small $\omega$
and therefore vanishes logarithmically at the Fermi surface.
The other two curves in Fig.~\ref{fig:compmar} include a DDW. The
dashed curve is based on a constant quasiparticle scattering rate of $\eta=0.1$
while the solid curve is for $0.1+0.5|\omega|$. 
The interband contribution manifests itself as a broad peak
centered around $\Omega\simeq 1.4t$, in the dashed curve,
which is also reasonably
well separated from the Drude contribution centered around $\Omega=0$.
As the impurity scattering is increased, the region
between the interband and the intraband fills in, and the interband peak
broadens and is reduced in height (not shown). In a sense, the
MFL result (black curve) can be understood as a case where
impurity scattering effectively increases with increasing frequency.
This has the outcome of spreading the interband spectral weight over a 
much larger frequency range reducing the prominence of the peak seen
in the dashed curve and shifting it to lower energy. 
Because of this shift, it is not possible to accurately determine
the size of the pseudogap from the effective position of the interband
transitions in the real part of the conductivity.
Nevertheless,
a peak still remains in the MFL case, although now it is not so well
separated from the intraband contribution at low energies and the 
$2|\mu|$ cutoff is no longer visible.

We also note from Eq.~(\ref{eq:resigma}), that for the 
DC conductivity at zero temperature, the 
electron spectral functions $A_{ij}(\bf{k},\omega)$ are all evaluated
at zero frequency. In this case, only the impurity self-energy remains, 
since $\Sigma_{MFL}(\omega=0)=0$ in Eq.~(\ref{eq:mflself}), therefore,
both the
impurity and the marginal Fermi liquid models will have the same DC
value because we have chosen the same $\eta$ and plasma frequency
in both cases.
However, for finite temperature, the marginal Fermi liquid 
DC conductivity will be reduced because of the inelastic 
scattering, a feature not present in the impurity case. 
When the DDW gap is included, spectral weight is transferred 
to the interband processes. In both cases of impurity only and
impurity plus MFL, this transfer of spectral weight is about 
$50\%$ for the parameters chosen, namely 
$\mu=-0.5t$ and $\Delta=0.8t$. This is consistent with the solid 
curves in Fig.~\ref{fig:transfer} which cross at this value of 
the chemical potential for $t'=0$. For the impurity case the value 
of the DC conductivity is reduced to $53\%$ of its original value 
while for the marginal Fermi liquid the percentage is $47\%$.

In Fig.~\ref{fig:lorenz}, we show our results 
for the optical scattering rate
obtained from the marginal Fermi conductivity of Fig.~\ref{fig:compmar}.
What is shown as the solid curve
is the usual $1/\tau_{op}(\omega)$ of 
Eq.~(\ref{eq:taudef}). It is this quantity that is used in the
experimental literature. However, it is useful to define a new
scattering rate denoted by $1/\tau^*_{op}(\omega)$ which is related to
$1/\tau_{op}(\omega)$ by a simple multiplicative constant.
For $1/\tau^*_{op}(\omega)$, what is used for forming the righthand
side of Eq.~(\ref{eq:taudef}) is the value of the DC conductivity
divided by $2\eta$, rather than $\Omega_p^2/4\pi$. For the impurity
case without the DDW, this would give $\Omega_p^2/4\pi$, as would also
be the case for the pure MFL at $T=0$. At $T\ne 0$, an inelastic contribution
to the scattering would also need to be added to $2\eta$. For the
case shown here, where $T=0.016t$, this is about a 10\% correction. As we
have seen, when the DDW is present, considerable spectral weight is transferred
from the intraband to the interband absorption, and so the Drude 
centered about
$\Omega=0$ is strongly depleted, and the DC conductivity is correspondingly
reduced. This means that $1/\tau^*_{op}(\omega)$ is about one half the value
of $1/\tau_{op}(\omega)$ for the case considered here when the DDW is present.
An important feature of $1/\tau^*_{op}(\omega)$ is that as $\omega\to 0$,
it agrees exactly with the residual scattering rate $2\eta$ with two
modifications. First, when quasiparticle scattering is present, the 
interband contribution can make a contribution to the DC conductivity,
and second, 
at finite temperature in the MFL, there can also be a small increase
in the DC scattering due to the inelastic scattering. Both these effects
can modify somewhat the $\omega\to 0$ of the $1/\tau^*_{op}(\omega)$.
Returning now to Fig.~\ref{fig:tau}, we can understand why, in the
lower temperature run, the DC limit of $1/\tau_{op}(\omega)$ is more than
twice the value of $2\eta$. It reduces with increasing temperature but
even for the solid curve at $T=0.95T^*$, it is still above $2\eta=0.1t$.
Returning to the solid curve of Fig.~\ref{fig:lorenz}, we note that its
DC value is also above $2\eta$ mainly due to the depletion of the intraband
contribution. At higher frequencies, it shows the interband peak slightly
below $\Omega/t=0.5$ and then takes on a quasilinear behaviour as expected
of the pure MFL, but with a slope which is considerably reduced over the
input scattering rate, shown as the dotted curve. The reduction in slope
can be traced to the presence of the interband transitions as will be made
clear in the following.

We now turn to the shift in the interband structure toward lower frequency 
seen in the scattering rate of Fig.~\ref{fig:lorenz} when compared to 
the structure in the conductivity of Fig.~\ref{fig:compmar}. 
This also holds for Fig.~\ref{fig:tau} 
when compared to Fig.~\ref{fig:sigmaevTemp}.
Some understanding of this can be obtained from the study of  
a simplified model. For convenience, we use a simplified form for the  
marginal Fermi liquid conductivity:\cite{tanner-timusk}
\begin{equation}
\sigma^{MFL}(\omega)
=\frac{(\Omega^{MFL}_p)^2}{4\pi}
\frac{1}{-i\omega(1-2\lambda \log |\omega/2\omega_c|) +
(2\eta+\pi\lambda \omega)}.
\label{eq:sigmalormfl}
\end{equation}
Note that for the conductivity, it is $2\eta+\pi\lambda|\omega|$
which enters as the damping rate. The constant impurity scattering rate
is twice its quasiparticle counterpart, but the MFL piece remains
unaltered. 
Eq.~(\ref{eq:sigmalormfl}) is a reasonably useful
 approximation to the more accurate calculation
that would arise from implementing the exact form for the normal state
conductivity:\cite{frank}
\begin{equation}
\sigma(\nu)
=\frac{\Omega_p^2}{4\pi}
\frac{i}{\nu}\int_0^\nu\, d\omega \frac{1}{\nu-\Sigma(\omega)-\Sigma(\nu-\omega)+i2\eta}
\label{eq:sigmanorm}
\end{equation}
with the marginal Fermi liquid self-energy given in Eq.~(\ref{eq:mflself}).
Adding to Eq.~(\ref{eq:sigmalormfl}) a form based on
a displaced Lorentzian oscillator in the dielectric function
of energy $\omega_E$, which is
used to model the interband contribution:\cite{ziman}
\begin{equation}
\sigma^{Lor}(\omega)
=\frac{(\Omega_p^{Lor})^2}{4\pi}\frac{-i\omega}{(\omega_E^2-
\omega^2)-i\omega\Gamma},
\label{eq:sigmalor}
\end{equation}
we form the optical scattering rate defined in Eq.~(\ref{eq:taudef})
for the combined system with one change. 
Instead of using the complete plasma frequency 
coming from interband and intraband combined, we use only 
the intraband contribution. This is done to define $1/\tau^*_{op}(\omega)$. 
We use Eq.~(\ref{eq:sigmalormfl}) and 
Eq.~(\ref{eq:sigmalor}) to model as well as possible the solid 
curve (MFL+DDW) in Fig.~\ref{fig:compmar}. The result is 
shown in Fig.~\ref{fig:lorenz} as the dashed line. The 
parameters used are $\lambda=1/(2\pi)$, $\eta=0.1t$, $\Gamma=0.9t$, 
$\omega_E=0.6t$ and $(\Omega_p^{Lor}/\Omega_p^{MFL})^2 =0.57$. 
With these parameters we can calculate the 
scattering rate $1/\tau^*(\Omega)$ for our model. 
It is given by the dash-dotted curve which can be compared with the 
dotted straight line equal to $0.2+0.5\Omega/t$.
These quantities agree at $\Omega=0$ as expected.
It is to be noted that this graph provides a good qualitative 
confirmation of the complete numerical results of 
Fig.~\ref{fig:lorenz} (solid curve). 
We note that $1/\tau^*(\Omega)$
differs from $1/\tau(\Omega)$ of formula Eq.~(\ref{eq:taudef})
mainly by a constant
numerical factor. We also note that the peak in the
scattering rate has shifted downward as compared with 
its position in the conductivity. This is clearly due to the 
particular combination of real and imaginary part of the 
conductivity defined in Eq.~(\ref{eq:taudef}). 
The reduced slope of the quasi-linear behaviour of the dash-dotted
curve at large $\Omega$ when compared with the pure MFL slope
is reproduced in our model (solid curve). It can now be seen that
it has its origin in the
fact that an interband contribution is present and is not due to
an alteration of the MFL representation of the remaining intraband
contribution (with the lorentzian term set to zero, we recover the
straight dotted line as expected).
We can conclude
from this analysis that, to good approximation, the complete conductivity
can be nicely modelled as a superposition of a MFL component 
of reduced spectral weight, but otherwise unaltered,
plus an interband component given
approximately by Eq.~(\ref{eq:sigmalor}). This reproduces the full
numerical calculation reasonably well, however, the original
pseudogap value and cutoff $2|\mu|$ could never be inferred from
fitting the experimental data to this model.

In principle, the formation of a DDW will also lead to changes in the 
self-energy of the charge carriers. For example in the DDW model the 
electronic density of states acquires additional energy dependence 
as seen in Fig.~\ref{fig:dos} which, in turn alters the quasiparticle 
scattering.\cite{hirschfeld,kimcarbotte} 
Here, we have modelled this scattering either with a constant 
elastic scattering rate or, to be more realistic, by a marginal 
Fermi liquid (MFL) contribution. This contribution to the inelastic 
scattering is directly responsible for the quasilinear dependence, 
in frequency 
$\omega$, of the optical scattering rate, seen in 
Fig.~\ref{fig:lorenz} at $\omega$ beyond the interband contribution. 
This quasilinear dependence is a hallmark of the cuprates 
and is central to the MFL phenomenology which was introduced to correlate
a large number of the normal state properties of the 
high $T_c$ oxides, which are anomalous. Unfortunately, there is no 
generally accepted underlying microscopic framework\cite{micro}
to the marginal Fermi liquid ansatz, which could be employed to introduced the 
necessary modifications to the self-energy that result 
from the DDW formation. Such considerations go beyond the scope 
of the present study.

The effects of energy dependence in the electronic density of 
states $\rho(\epsilon)$ on the self-energy and on the conductivity 
have recently been studied\cite{meirong} for elastic 
impurity scattering and are well understood. For example, 
a depression in $\rho(\epsilon)$ below its constant background 
value around the Fermi energy, leads directly
to a corresponding reduction in the optical scattering rate 
in this same frequency region. In a non-self-consistent approach, 
the constant quasiparticle scattering rate of the flat $\rho(\epsilon)$
case, is simply multiplied by $\rho(\epsilon)$ to lowest
order. In our case, this would be the $\rho(\epsilon)$ of 
the DDW as shown in Fig.~\ref{fig:dos}. This would correspondingly 
reduce the scattering rate of Fig.~\ref{fig:lorenz} at small 
$\omega$. However, it was found in reference\cite{meirong} that such 
a procedure overestimates the effect of $\rho(\epsilon)$ on the 
quasiparticle as well as on the optical scattering rate. In 
a self-consistent approach, the impurities and, even more
importantly, the inelastic scattering  of the MFL smear out the 
structures in $\rho(\epsilon)$ which are due to the opening of the DDW gap. 
Nevertheless, a reduction of $\tau^{-1}(\Omega)$ below the 
value seen in Fig~\ref{fig:lorenz} is expected for frequencies 
below the gap and would be associated with self-energy corrections.

Finally, we discuss the issue of comparison with experiment.
This model predicts the appearance of an interband contribution in
the pseudogap phase and an accompanying peak in the optical
scattering rate. This interband contribution should have a feature
at lower energy corresponding to a lower cutoff set by $2|\mu|$
and a higher cutoff of about $2\Delta$. The lower feature should
shift to higher energy with increased doping, however,
from experiment the gap decreases with doping and hence the 
upper edge in the conductivity should decrease with
increased doping.
This has not been seen in
experiments on the high $T_c$ cuprates. Examples of
optical experiments examining
the underdoped state of various cuprates are given in Refs.~\cite{puchkov,
puchkovprl,takenaka} for YBCO, Bi2212, and LSCO materials. There is no
evidence for an interband feature with energy scales as described here
and the optical scattering rate\cite{puchkov} does not show a peak. The
most up-to-date review on optical conductivity experiments in high
$T_c$ cuprates\cite{basov} does not give further evidence that
could support this model. Indeed the DDW model has been controversial in
its applicability to experiments in the cuprates. Some criticisms have
involved the variance of the model with regard to photoemission\cite{zxshen} 
and 
tunneling\cite{kimcar2}, 
but some of these issues have been possibly rectified by
further calculation\cite{chakraphoto,greco2}. 
Here we find that the experimental evidence from
optical conductivity, at this
time, does not support the DDW model for the cuprates. However, this
model may find application in other systems in the future.

\section{Conclusions}
We have calculated the optical conductivity and the Raman response for 
a DDW system. There is a cutoff 
at low frequencies in the interband response 
for both the optical conductivity and the Raman response.
This cutoff is at $2|\mu|$ if there is no imperfect nesting
term in the dispersion relation.
The imperfect nesting adds 
directly to the chemical potential and shifts 
the cutoff to lower frequency but does not eliminate it. 
Including scattering makes the cutoff harder to discern.

As the chemical potential 
is changed we find a readjustment of the spectral weight 
between the inter- 
and intraband conductivities. That part of 
the interband contribution, ranging from 
frequencies zero 
to $2|\mu|$,
is transferred at zero temperature 
to the Drude weight as the chemical potential 
is increased with increasing doping. 
However, contrary to a first expectation there is little loss in the total
spectral weight with the opening of the pseudogap. This 
result agrees with 
experiments.\cite{bontemps} 

As mentioned above, the $2|\mu|$ 
cutoff also appears in the electronic 
Raman scattering. The best channels to observe the cutoff at $2|\mu|$ are 
$B_{2g}$ and $A_{1g}$. The $B_{1g}$ channel projects out the 
antinodal region where the gap is biggest, and 
the low intensity of the Raman scattering at small $\omega$, 
where it goes like $(\omega/\Delta)^3$, may make
the cutoff difficult to see.

The behaviour with temperature of the optical conductivity has two 
characteristics: 1) the expected shift of the $2\Delta$ feature 
to lower frequencies, for a mean field theory with 
order parameter vanishing at $T^*$, and 2) the 
transfer of spectral weight from interband to intraband as 
the interband contribution collapses. There is also a filling 
of the 
depression between low frequencies and $2|\mu|$.

An important result of this paper is that there is a 
peak in the 
real part of the optical conductivity 
in the region 
$\Omega_c<\Omega<2\Delta$ that comes from the 
interband conductivity with $\Omega_c=2|\mu|$ for $t'=0$.
This peak is robust against different dispersion relations including 
second-nearest-neighbours. It is reduced by the impurity scattering
and by the inelastic
scattering, which can also shift it.
It is washed out with increasing temperature
vanishing completely at the pseudogap critical temperature.
Similar structure is found in the optical scattering rate, but it
is shifted relative to that in the optical conductivity.
With strong inelastic scattering, as may be found in the MFL model,
the $2|\mu|$ cutoff and $2\Delta$ features become more difficult
to determine directly.
To the best of our knowledge, this interband structure has not yet
been identified  in
experiments on cuprates.

\begin{acknowledgments}
EJN acknowledges funding from NSERC and the
Government of Ontario
(Premier's Research Excellence Award), and the University of Guelph.
JPC acknowledges support from NSERC and the CIAR. We 
thank L. Benfatto, W. Kim, and D. Basov
for helpful discussions.
\end{acknowledgments}



\begin{thebibliography}{99}
\bibitem{timuskpseudo} T. Timusk and B. Statt, Rep. Prog. Phys.
{\bf 62}, 61 (1999) and references therein.
\bibitem{ding} H. Ding, T. Yokoya, J. C. Campuzano, T. Takahashi, 
M. Randeria, M.R. Norman, T. Mochiku, K. Kadowaki, and J. Giapintzakis, 
Nature (London) {\bf 382}, 51 (1996).
\bibitem{shen} A.G. Loeser, Z.-X. Shen, D.S. Dessau, D.S. Marshall, 
C.H. Park, P. Fournier, and A. Kapitulnik, Science {\bf 273}, 325 (1996).
\bibitem{renner} Ch. Renner, B. Revaz, J.-Y. Genoud, K. Kadowaki, 
and O. Fischer, Phys. Rev. Lett. {\bf 80}, 149 (1998).
\bibitem{kivelson} V. J. Emery and S.A. Kivelson, Nature (London) 
{\bf 374}, 134 (1995).
\bibitem{emery} E. Carlson, V.J. Emery, S.A. Kivelson, and 
D. Orgad in {\it The Physics of Conventional and Unconventional 
Superconductors} edited by K.H. Bennemann and J.B. Ketterson 
(Springer-Verlag, New York, 2004).
\bibitem{chen} Q.J. Chen, I. Kosztin, B. Jank\'o, and K. Levin, 
Phys. Rev. Lett. {\bf 81}, 4708 (1998).
\bibitem{chakra} S. Chakravarty, R.B. Laughlin, D.K. Morr,
and C. Nayak, Phys. Rev. B {\bf 63}, 094503 (2001).
\bibitem{chakra2} S. Chakravarty, H.-Y. Kee, and C. Nayak, 
Int. J. Mod. Phys. B {\bf 15}, 2901 (2001).
\bibitem{wang} Q. -H. Wang, J.H. Han, and D -H. Lee, Phys. Rev. 
Lett. {\bf 87}, 077004 (2001).
\bibitem{carbotte} J. -X. Zhu,W. Kim, C. S. Ting, and J. P. Carbotte,
Phys. Rev. Lett. {\bf 87}, 197001 (2001).
\bibitem{nayak} X. Yang and C. Nayak, Phys. Rev. B {\bf 65}, 064523 
(2002).
\bibitem{maki} B. D\'ora, A. Virosztek, and K. Maki, Phys. Rev. B
{\bf 65}, 155119 (2002). K. Maki, B. D\'ora, M. Kartsovnik, A.
Virosztek,B. Korin-Hamzic,and M. Basletic, Phys. Rev. Lett.
{\bf 90}, 256402 (2003).
\bibitem{dora} B. D\'ora, K. Maki, and A. Virosztek, 
Mod. Phys. Lett. B. {\bf 18}, 327 (2004).
\bibitem{chakraphoto} S. Chakravarty, C. Nayak, and S. Tewari 
Phys. Rev. B {\bf 68}, 100504 (2003).
\bibitem{schachi} I. Sch\"urrer,
E. Schachinger and  J.P. Carbotte, Physica C  {\bf 303}, 287 (1998). 
\bibitem{molegraaf}H.J.A. Molegraaf, C. Presura, D. van der Marel, 
P.H. Kes, and M. Li, Science {\bf 295} 2239 (2002).
\bibitem{bontemps} A. F. Santander-Syro, R.P.S.M. Lobo, N. Bontemps,
Z. Konstantinovic, Z. Li,and H. Raffy, Phys. Rev. Lett. {\bf 88} 97005 (2002).
\bibitem{benfatto} L. Benfatto, S.G. Sharapov and H. Beck, 
Eur. Phys. J. {\bf 39}, 469 (2004).
\bibitem{aristov} D.N. Aristov and R. Zeyher, cond-mat/0406419
\bibitem{benfatto2} L. Benfatto, S.G. Sharapov, N. Andrenacci, and H. Beck, 
Phys. Rev. B {\bf 71}, 104511 (2005).
\bibitem{shastry} B.S. Shastry and B.I. Shraiman, 
Int.J.Mod.Phys. B {\bf 5} 365 (1991). D. Branch and J.P. Carbotte, Jour. 
Supercon. {\bf 13} 535 (2000).
\bibitem{dwayne} D. Branch, MSc Thesis McMaster University, (1995).
\bibitem{t} Strictly speaking, 
$\frac{\sinh(\beta\Omega/2)}{\cosh(\beta\mu)+\cosh(\beta\Omega/2)}$ 
is a universal function 
only for $t'=0$. However, since we already know that the effect of $t'$ 
is to shift 
the cutoff at $2|\mu|$, we still use this universal function in 
the continuum limit. 
\bibitem{greco} R. Zeyher and A. Greco, Phys. Rev. Lett. {\bf 89},
177004 (2002). 
\bibitem{puchkov} A.V. Puchkov, D.N. Basov, and T. Timusk,
J. Phys.: Condens. Matter {\bf 8}, 10049 (1996).
\bibitem{tanner-timusk} D.B. Tanner and T. Timusk, in \emph{The Physical
Properties of High Temperature Superconductors III},D.M. Ginsberg, Ed.
(World Scientific, Singapore, 1992),pp. 363-469.
\bibitem{timuskCo} J. Hwang, J. Yang, T. Timusk, and F.C. Chou,
cond-mat/0405200.
\bibitem{mfl} C. M. Varma, P. B. Littlewood, S. Schmitt-Rink, E. 
Abrahams and A. E. Ruckenstein, Phys. Rev. Lett. {\bf 63}, 1996 (1989).
\bibitem{varma} C. M. Varma, Phys. Rev. B {\bf 55}, 14554 (1997).
\bibitem{frank} F. Marsiglio, T. Startseva, and J.P. Carbotte, Phys. Lett.
A {\bf 245}, 172 (1998).
\bibitem{ziman} See for instance, J.M. Ziman, Principles of the
Theory of Solids (Cambridge University Press, 1965).
\bibitem{hirschfeld} P.J. Hirschfeld, P. W\"olfle, and D. Einzel, 
Phys. Rev. B {\bf 37}, 83 (1988); P.J. Hirschfeld, W.O. Putikka, 
and D.J. Scalapino, Phys. Rev. B {\bf 50} 10250 (1994).
\bibitem{kimcarbotte} W. Kim and J.P. Carbotte, Phys. Rev. B {\bf 66},
033104 (2002).
\bibitem{micro} Theories that reproduce the marginal Fermi liquid 
behaviour for a specific region of the $x-T$ phase diagram 
of the cuprates are the nested Fermi liquid, A. Virosztek and J. Ruvalds, 
Phys. Rev. B {\bf 42},4064 (1990)
and the so-called Van Hove scenario, J. Gonz\'alez, F. Guinea, and M.A.H. 
Vozmediano, Nucl. Phys. B {\bf 485}, 694 (1997).  
\bibitem{meirong} M.-R. Li and J.P. Carbotte, Phys. Rev. B {\bf 60},
155114 (2002).
\bibitem{puchkovprl} A.V. Puchkov, P. Fournier, T. Timusk, and N.N. Kolesnikov,
Phys. Rev. Lett. {\bf 77}, 1853 (1996).
\bibitem{takenaka} K. Takenaka, J. Nohara, R. Shiozaki, and S. Sugai,
Phys. Rev. B {\bf 68}, 134501 (2003).
\bibitem{basov} D.N. Basov and T. Timusk, Rev. Mod. Phys., in press.
\bibitem{zxshen} A. Damascelli, Z. Hussain, and Z-X. Shen, Rev. Mod. Phys.
{\bf 75}, 473 (2003).
\bibitem{kimcar2} W. Kim, J.-X. Zhu, J.P. Carbotte, and C.S. Ting,
Phys. Rev. B {\bf 65}, 064502 (2002).
\bibitem{greco2} A. Greco and R. Zeyher, Phys. Rev. B {\bf 70}, 024518 (2004).
\end{thebibliography}
\end{document}